%% file: main.tex
\documentclass[lettersize,journal]{IEEEtran}
\usepackage{amsmath,amsfonts}
\usepackage{algorithmic}
\usepackage{algorithm}
\usepackage{array}
\usepackage[caption=false,font=normalsize,labelfont=sf,textfont=sf]{subfig}
\usepackage{textcomp}
\usepackage{stfloats}
\usepackage{url}
\usepackage{verbatim}
\usepackage{graphicx}
\usepackage{cite}
\usepackage{hyperref}
\usepackage{cleveref}
\usepackage{xspace,xpunctuate}
\usepackage{enumitem}
\newcommand{\eg}{\textit{e.g.},\xspace}
\newcommand{\etal}{\xspace\textit{et al.}\xspace}

\begin{document}

\title{DashChat: Interactive Authoring of Performance Dashboard Design Prototypes through Conversation with LLM-Powered Agents}

\author{Ziyue Lin, Siqi Shen, Honghui Mei, Wanchen Liu, Chengye Xin, Wenzhuo Dai, Siming Chen, Xiao Wen, and Xingyu Lan
\thanks{Ziyue Lin, Siming Chen, Xingyu Lan are with Fudan University, China. (e-mail: ziyuelin917@gmail.com; simingchen@fudan.edu.cn; xingyulan96@gmail.com). }
\thanks{Siqi Shen, Honghui Mei, Wanchen Liu, Chengye Xin, Wenzhuo Dai, Xiao Wen are with DataV Lab, Alibaba Group, Hangzhou, China. (e-mail: shensiqi.ssq@alibaba-inc.com; honghui.mhh@alibaba-inc.com; liuwanchen.lwc@alibaba-inc.com; mier.xcy@alibaba-inc.com; basi.dwz@alibaba-inc.com; ninglang.wx@alibaba-inc.com).}
\thanks{Ziyue Lin and Siqi Shen contributed equally to this research. Xiao Wen and Xingyu Lan are the corresponding authors.}}

\markboth{Journal of \LaTeX\ Class Files,~Vol.~14, No.~8, August~2021}%
{Shell \MakeLowercase{\textit{et al.}}: A Sample Article Using IEEEtran.cls for IEEE Journals}


\maketitle

\begin{abstract}
\input{sections/0abstract}
\end{abstract}


\input{sections/1introduction}
\input{sections/2related_work}
\input{sections/3overview}
\input{sections/3.5design_pattern}
\input{sections/4system}

\input{sections/4.5interface}
\input{sections/5evaluation}
\input{sections/6discussion}

\bibliographystyle{IEEEtran}
\bibliography{reference}

\begin{IEEEbiography}[{\includegraphics[width=1in,height=1.25in,clip,keepaspectratio]{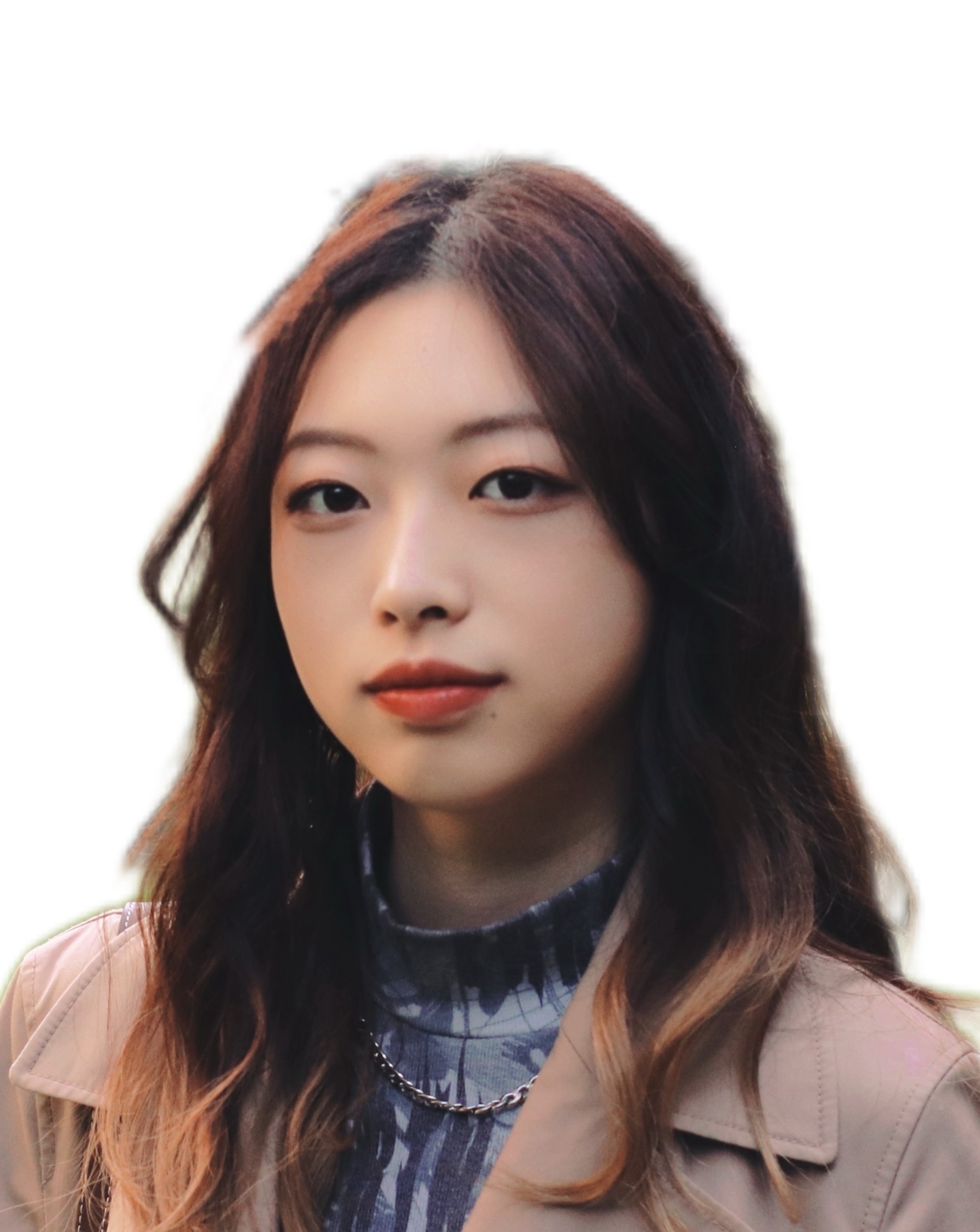}}]{Ziyue Lin} received the BS degree from Fudan University. She is currently working toward the PhD degree with the School of Data Science, Fudan University. Her primary research interests include visualization, human-computer interaction, artificial intelligence, and agent-based modeling.
\end{IEEEbiography}

\begin{IEEEbiography}
[{\includegraphics[width=1in,height=1.25in,clip,keepaspectratio]{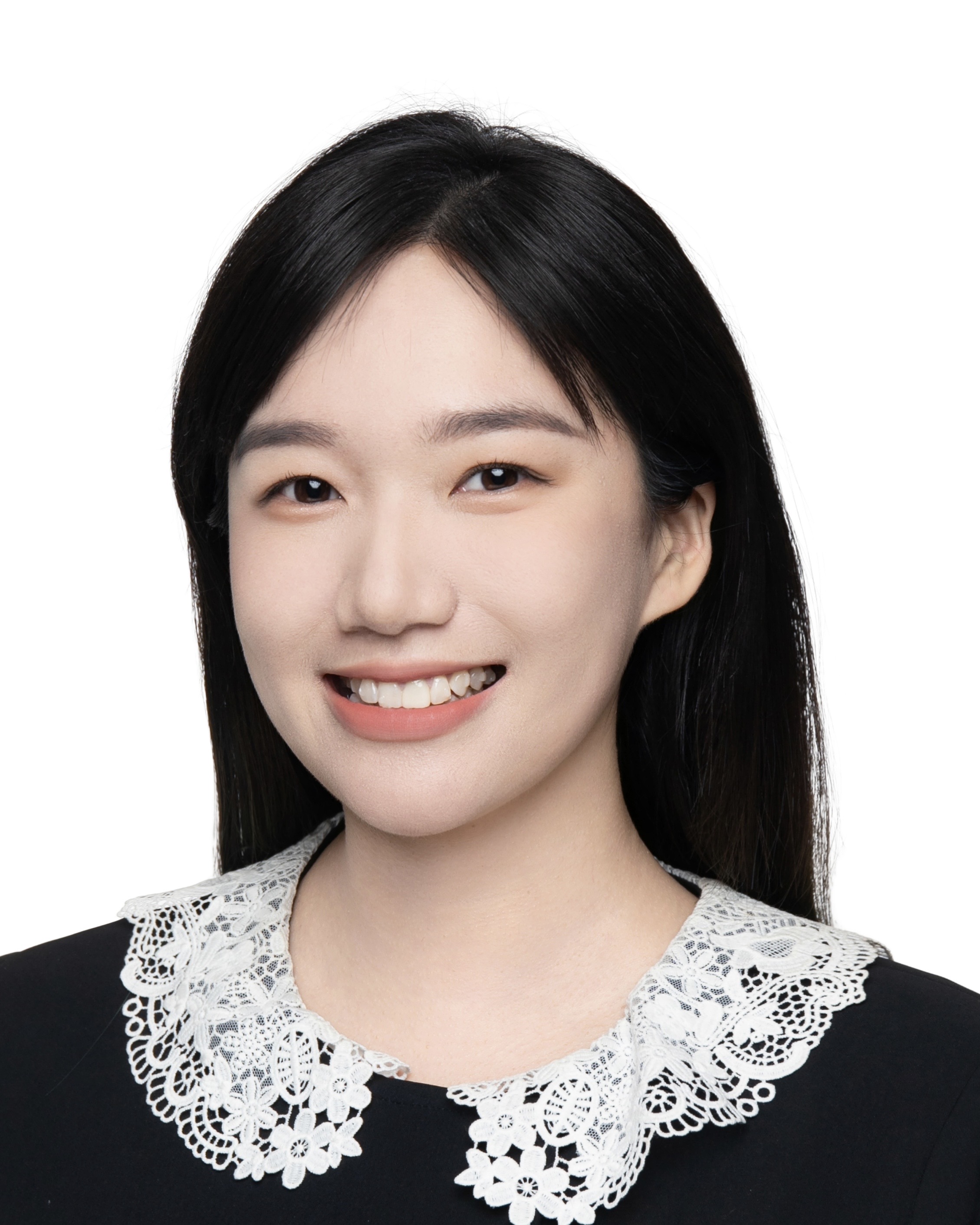}}]{Siqi Shen} received the master’s degree from Fudan University, China, in 2022. She is currently working with Alibaba Cloud Computing, Hangzhou, China. Her research interests include information visualization, human-computer interaction and artificial intelligence.
\end{IEEEbiography}

\begin{IEEEbiography}
[{\includegraphics[width=1in,height=1.25in,clip,keepaspectratio]{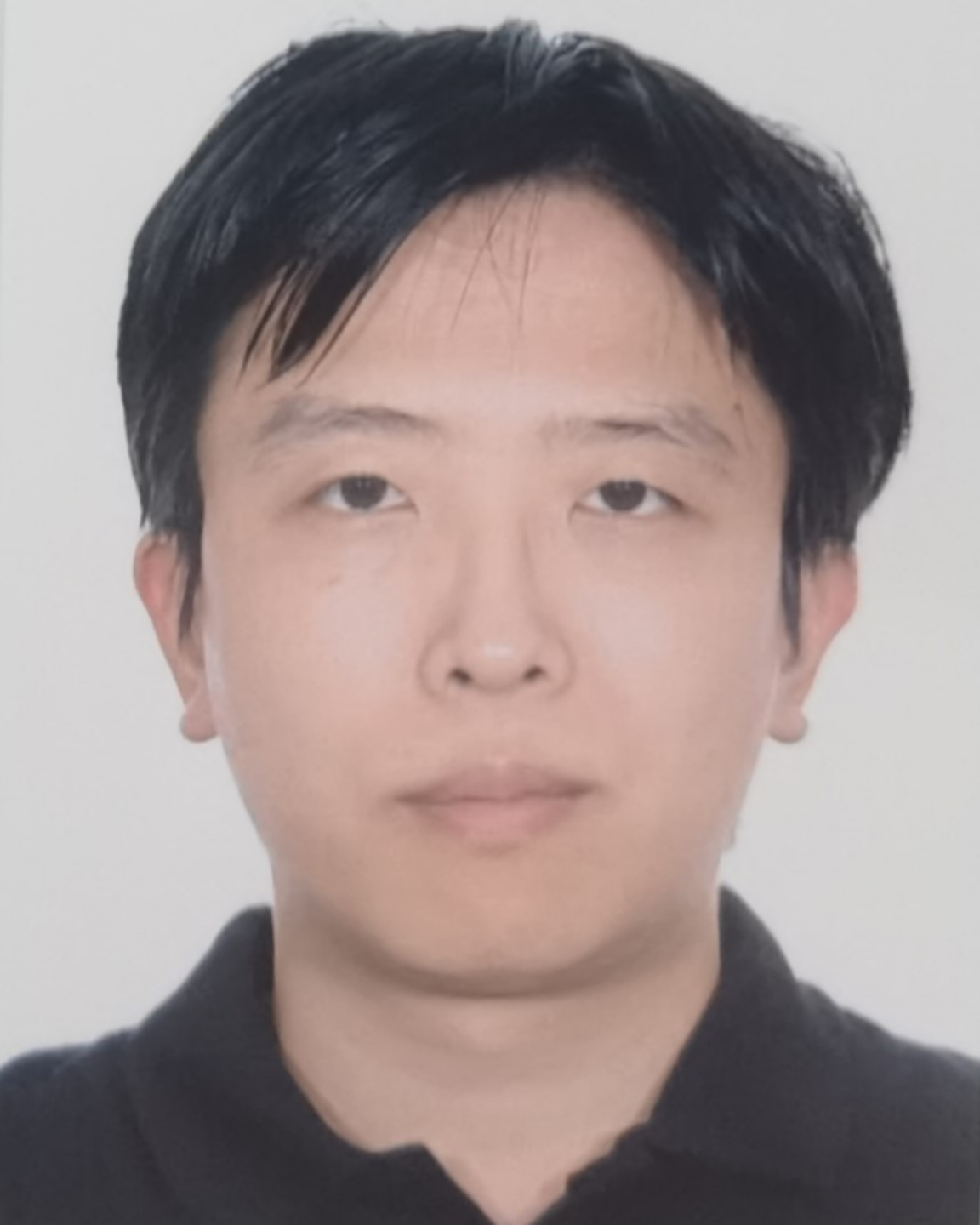}}]{Honghui Mei} received the PhD degree from the State Key Lab of CAD \& CG, Zhejiang University, China, in 2019. He is currently working with Alibaba Cloud Computing, Hangzhou, China. His research interests include information visualization and visual analytics.
\end{IEEEbiography}

\begin{IEEEbiography}
[{\includegraphics[width=1in,height=1.25in,clip,keepaspectratio]{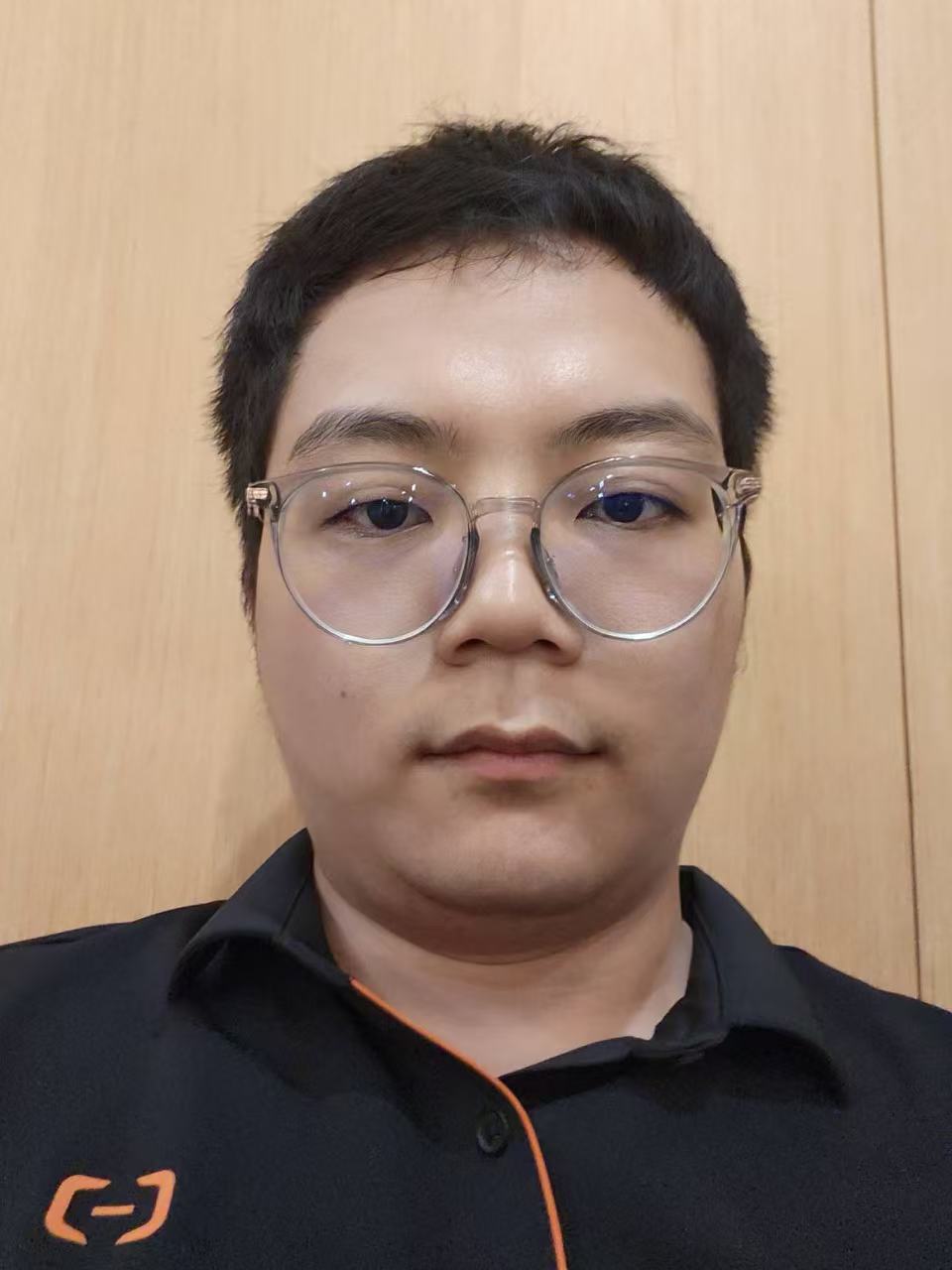}}]{Wanchen Liu} received the BS degree from Hangzhou Dianzi University, China, in 2022. He is currently working with Alibaba Cloud Computing, Hangzhou, China. His primary research interests include Human-AI Collaboration and Multi-Agent Systems.
\end{IEEEbiography}

\begin{IEEEbiography}
[{\includegraphics[width=1in,height=1.25in,clip,keepaspectratio]{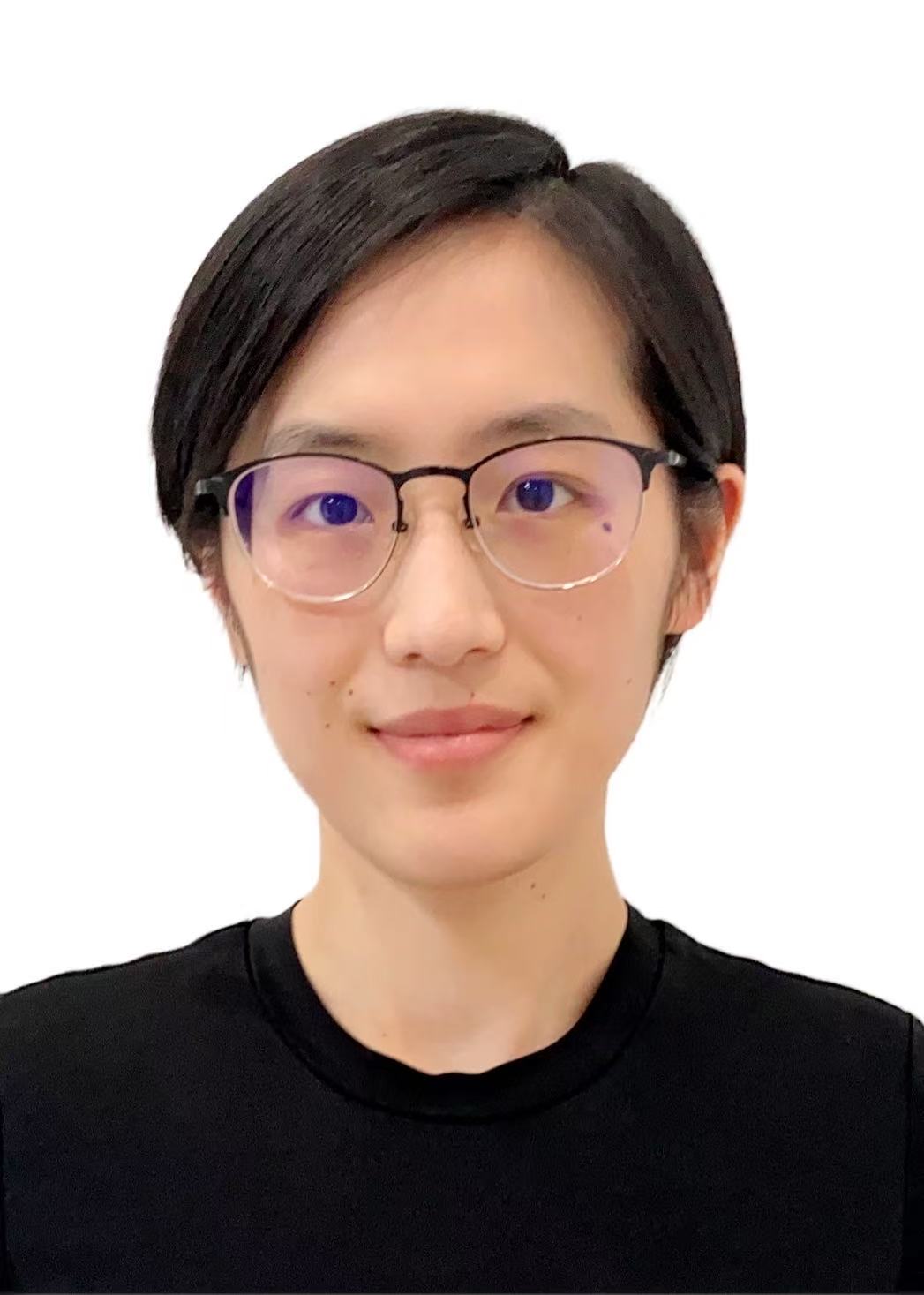}}]{Chengye Xin} received the BS degree from the Dalian University of Technology, Dalian, China. She is currently a senior manager at Alibaba Cloud Computing. Her research interests include information visualization and computing infrastructures.
\end{IEEEbiography}

\begin{IEEEbiography}
[{\includegraphics[width=1in,height=1.25in,clip,keepaspectratio]{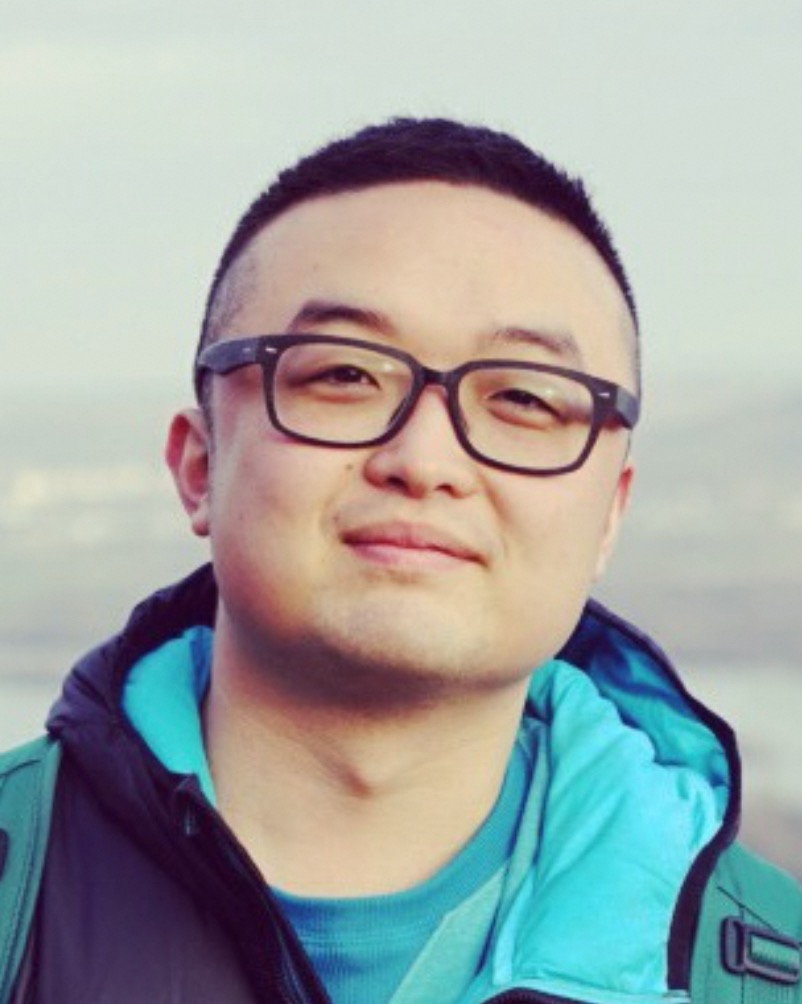}}]{Wenzhuo Dai} received the BS degree from Anhui Polytechnic University, Wuhu, China. He is currently a senior manager at Alibaba Cloud Computing. His research interests include computing infrastructures and database systems.
\end{IEEEbiography}

\begin{IEEEbiography}
[{\includegraphics[width=1in,height=1.25in,clip,keepaspectratio]{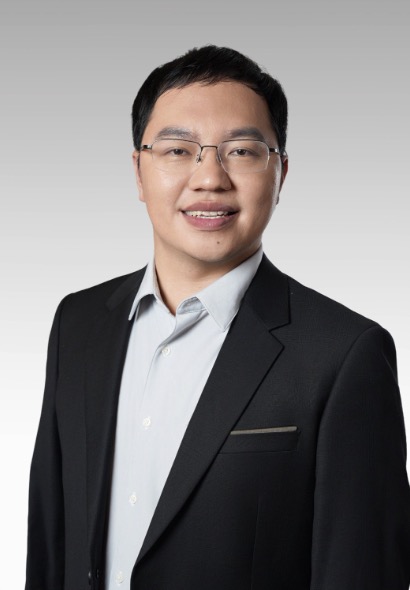}}]{Siming Chen} is an Associate Professor at Fudan University’s School of Data Science. Previously a Research Scientist at Fraunhofer IAIS, he earned his Ph.D. from Peking University. His research focuses on visualization and visual analytics, emphasizing Human-AI Collaboration. With over 100 publications, including 40+ in top venues like IEEE VIS and ACM CHI, he has received 10+ best paper and honorable mention awards.
\end{IEEEbiography}

\begin{IEEEbiography}
[{\includegraphics[width=1in,height=1.25in,clip,keepaspectratio]{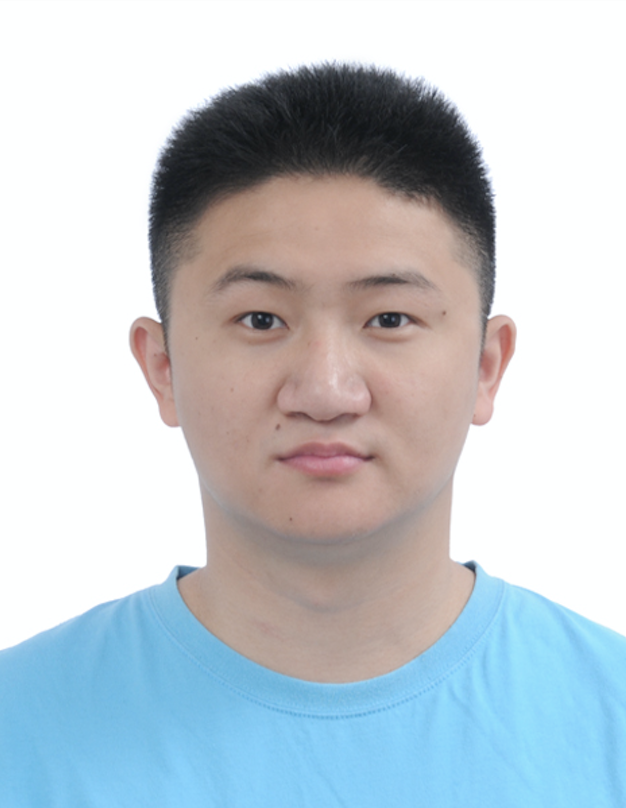}}]{Xiao Wen} received the BS degree from Zhejiang A\&F University, Hangzhou, China. He is currently a senior manager at Alibaba Cloud Computing. His research interests include information visualization and computing infrastructures.
\end{IEEEbiography}

\begin{IEEEbiography}
[{\includegraphics[width=1in,height=1.25in,clip,keepaspectratio]{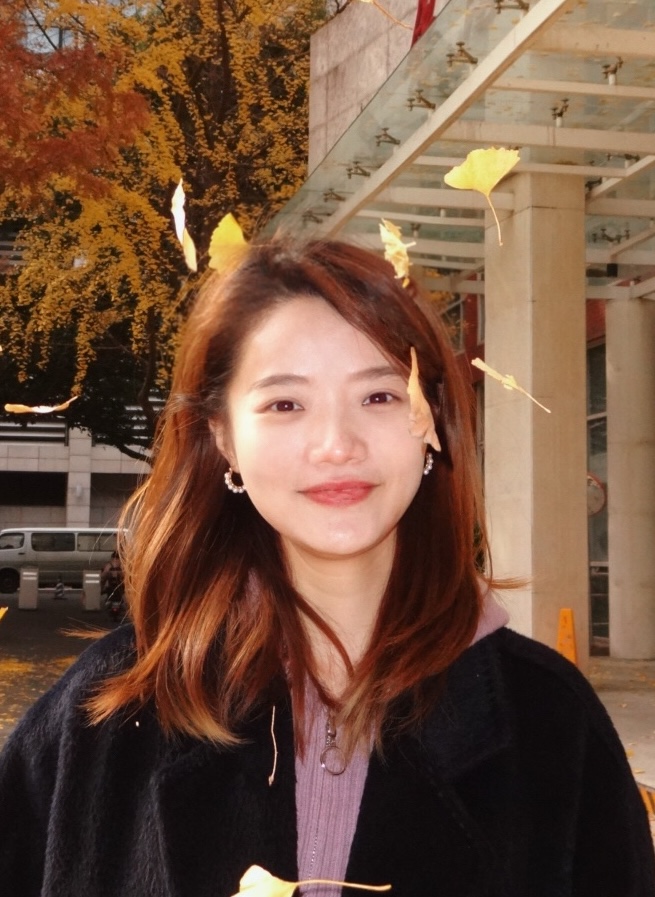}}]{Xingyu Lan} is an Assistant Professor at the School of Journalism, Fudan University, Shanghai, China. She earned her Ph.D. from Tongji University. Her research focuses on data storytelling, information design, user experience, and human-AI interaction. She received awards such as IEEE VIS Best Paper Award and Honorable Mention Award.
\end{IEEEbiography}
\vfill

\end{document}

%% file: sections/0abstract.tex
Performance dashboards are dashboards designed for and deployed within industrial settings (\eg enterprises, government agencies) to showcase and monitor their operational performance. They have evolved into an important and well-commercialized format for data visualization.
In practice, the ideation and negotiation phases demand rapid prototyping and iteration to align with evolving client needs.
However, existing tools compel designers to compromise either on iteration speed or on the meticulous handling of visual complexities.
To address these gaps, we introduce DashChat for generating performance dashboard prototypes.
Collaborating with industry experts, we derived the design requirements and analyzed 114 dashboards to extract common design patterns.
Informed by the findings, our solution integrates a chat interface with an LLM-driven multi-agent pipeline, translating textual requirements into prototypes.
We evaluated the system by comparing it with a baseline, demonstrating its effectiveness in facilitating the prototyping process while ensuring design quality.

%% file: sections/1introduction.tex
\section{Introduction}
\label{sec:introduction}
\begin{figure}[ht]
\centering
\includegraphics[width=\linewidth]{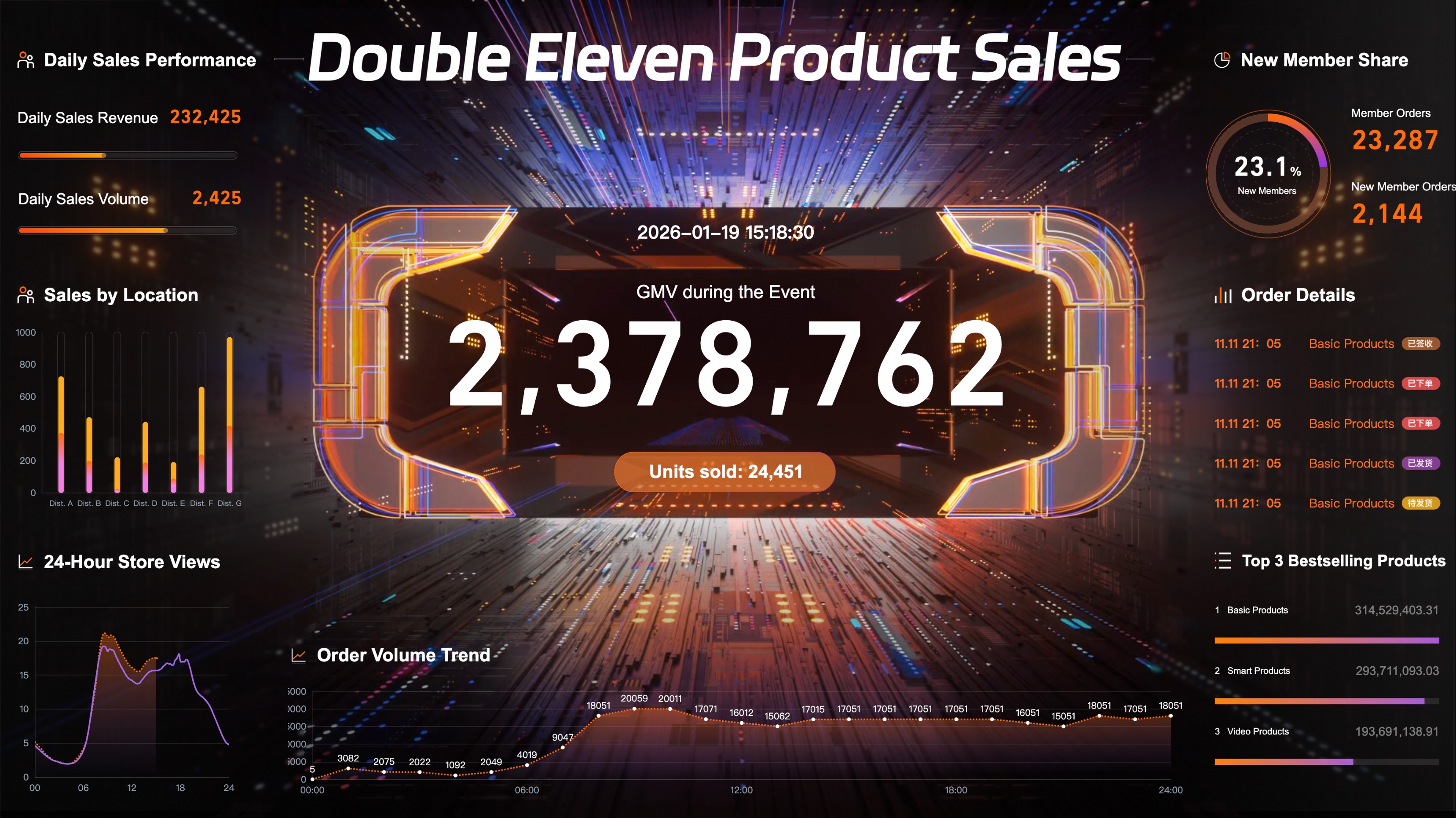}
\caption{Examples of Performance Dashboards: Double Eleven Product Sales Monitoring Dashboard.}
\label{fig:dashboardExamples}
\end{figure}

\IEEEPARstart{P}{erformance} dashboards are visualization systems designed for organizations such as enterprises and government agencies to measure and monitor their operational performance effectively~\cite{https://doi.org/10.1111/cgf.15001,8358558,eckerson2010performance}.
Often deployed in command centers and public spaces, performance dashboards have become increasingly critical in facilitating data communication and supporting informed decision-making across diverse domains~\cite{Mei:2020:DataV,kruglov2021tailored}.
Unlike dashboards designed to support exploratory analysis by data professionals, a performance dashboard is a full-fledged business information system built upon business intelligence and data integration infrastructures, which provides timely and high-level visibility into key performance indicators~\cite{8358558}.
An exemplary case of a performance dashboard is ``\textit{Alibaba Double Eleven Product Sales Monitoring Dashboard}'' (Fig.~\ref{fig:dashboardExamples}). 
The dashboard was designed for real-time sales tracking during the annual shopping extravaganza, enabling businesses to monitor key metrics like total sales, regional distribution, and top products through an intuitive interface.
Alibaba's success in the digital economy is made visible through the dashboard.
More than a promotional campaign, Double Eleven is setting benchmarks for online retail and reshaping contemporary shopping culture.
The examples demonstrate the transformative role of performance dashboards in addressing both commercial and societal needs.

However, based on our in-depth discussions with professional designers, performance dashboard design is a complex task, requiring the careful manipulation and coordination of various elements, including data visualizations, embellishments, layout structures, and animations~\cite{kruglov2021tailored, Sarikaya:2019:What, Bach:2023:Dashboard}.
Consequently, the design process often demands substantial expertise, including visual encoding, graphical design, and aesthetic considerations.
Additionally, in real-world industrial settings, designers are often required to rapidly produce design prototypes during early-stage collaboration and negotiation with clients.
It helps save costs before formal implementation and addresses potential data availability and security limitations. 
As a result, designers frequently engage in prototype design without access to the actual datasets, relying instead on abstract requirements or high-level data descriptions.
Although existing works have explored interactive authoring and recommendation tools for multi-view dashboards~\cite{10891199,10556805}, these methods are mostly not tailored to performance dashboard prototype generation.
MultiVision~\cite{Wu:2022:MultiVision} and DashBot~\cite{Deng:2023:DashBot} are data-centric, leveraging data characteristics and machine learning techniques to recommend charts tailored to user needs.
They have primarily focused on data mining and analysis, with relatively little attention to visual design, such as layout composition.
Rather than supporting analytical exploration, performance dashboards serve primarily as communication platforms for operational performance.
Meanwhile, design tools like Tableau~\cite{Tableau:2023:Learn} and Flourish~\cite{Flourish:2023:Flourish} are not tailored for performance dashboards and require substantial manual effort to create them.
In industrial settings, clients often expect an efficient creation process and deployment-ready dashboard prototypes.
While design ideation systems such as LADV~\cite{Ma:2021:LADV} have enabled convenient chart generation directly from visual inputs like hand-drawn sketches, there remains limited exploration into authoring tools that support dashboard prototype construction purely through natural language instructions.
Motivated by the limitations, we aim to develop an authoring tool to generate performance dashboard prototypes.
There are two major challenges to overcome.
First, dashboards are applied across various scenarios and exhibit substantial internal diversity.
For instance, dashboards designed for data analysis differ significantly from those intended for external presentation.
Despite current researches characterize design goals~\cite{Sarikaya:2019:What}, data presentations~\cite{Bach:2023:Dashboard} and layout arrangement~\cite{Chen:2021:Composition,Bach:2023:Dashboard} of general dashboard designs.
There is no systematic design guideline tailored explicitly for performance dashboards that focus on data external presentation.
The second challenge is to generate prototypes from vague, high-level text without specific data.
In real-world industrial scenarios, designers often provide abstract descriptions such as ``Global e-commerce sales dashboard showing country-level sales, daily sales by category, and quarterly net profit.'' without a dataset, making it difficult to determine appropriate visualization types and layout structures.
Accurately translating the unstructured text into compliant prototypes requires dedicated contextual inference and intelligent decision-making abilities, which impose high demands on our algorithm.

To address the challenges, we propose ${DashChat}$, an interactive system that leverages large language models~(LLMs) to generate performance dashboard design prototypes from natural language automatically.
For the first challenge, we collected 114 high-quality performance dashboards from various domains and summarized a design space that characterizes their view types, view layouts, analysis tasks, and embellishments.
The analysis results are then injected into LLMs as a reference during generation.
Second, we built a multi-agent pipeline powered by LLMs to generate effective and aesthetic prototypes.
We leverage LLMs' powerful language comprehension to develop intelligent agents with decision-making and diverse tool-use functionalities.
Additionally, due to the high computational cost of LLMs, our pipeline employs a parallel framework to enhance efficiency.
Based on the pipeline, we developed an interface consisting of the Chat Panel and History Panel, which supports users in authoring performance dashboard prototypes through conversations with agents and reviewing previous specifications.
It also assists users in clarifying their prototype requirements by offering multiple potential options, while supporting an iterative trial-and-error process informed by user feedback.
In conclusion, the key contributions of our work include: 
\begin{itemize}[leftmargin=*,topsep=0pt]
    \item A performance dashboard design space identified through an empirical analysis of 114 high-quality performance dashboards, which establishes a structured knowledge base to steer prototype generation. 
    \item A multi-agent automatic dashboard prototype generation pipeline for realizing an interactive system \textbf{DashChat}, that enables designers to construct effective and aesthetic performance dashboard prototypes through conversational interaction. It demonstrates the potential of LLMs to support real-world design workflows by translating natural language input into visual encodings, reducing the entry barrier for non-technical stakeholders.
    \item A quantitative study showing the effectiveness of our multi-agent generation pipeline and a user study to verify the usefulness and efficiency of our system.
\end{itemize}

%% file: sections/2related_work.tex
\section{Related Work}
\label{sec:related-work}
We review prior research on dashboard design patterns and guidelines, automated dashboard authoring tools, and LLM-powered agents and visualization design ideation.

\subsection{Dashboard Design Patterns and Guidelines}
The dashboard is a visual display consolidated within a single screen, designed to present the most critical information for supporting one or more data-driven insights~\cite{10.5555/1206491, Khan:2022:Rapid}.
Few\etal~\cite{few:2007:dashboard} argued that it can be applied both for monitoring in static forms and for analytical tasks when dynamic and interactive.
Researchers have agreed on some design guidelines:~(i) Avoiding visual clutters~\cite{Ogan:2012:Review, few:2007:dashboard};~(ii) Selecting indicators with care and avoiding too much data~\cite{Rahman:2017:Review, Janes:2013:Effective};~(iii) Aligning with user workflows~\cite{Faiola:2015:Supporting};~(iv) Providing consistency, interaction affordances and manage complexity~\cite{Sarikaya:2019:What};~(v) Organizing charts symmetrically and according to time~\cite{Bernard:2019:Using};~(vi) Balancing the tradeoff between detail density in a single view and the need for data access and exploration through charts~\cite{Bach:2023:Dashboard}.
Researchers also proposed common design patterns as useful speculative tools in the design process.
Sarikaya\etal~\cite{Sarikaya:2019:What} raised some common dashboard designs but mainly concluded features from multiple views of infographics and other storytelling genres. 
Bach\etal~\cite{Bach:2023:Dashboard} proposed a widely applicable dashboard pattern based on a contemporary corpus, dividing it into content and composition.
The content pattern includes data, meta information, and visual representations, while the composition pattern covers layout, screenspace, structure, interaction, and color.
Similarly, Chen\etal~\cite{Chen:2021:Composition} studied real-world dashboard designs and proposed two fundamental measures, composition that quantifies view types and configuration which characterizes view layouts. 
Meanwhile, many regarded dashboard design as a layout design problem and proposed rule-based layout approaches based on design principles~\cite{xu:2014:global, Donovan:2014:Learning, Sarikaya:2019:What, Zheng:2021:Sketchnote}. 
On the contrary, others explored exemplar-based approaches based on design instances. 
O'Donovan\etal~\cite{Donovan:2014:Learning}, Zheng\etal~\cite{Zheng:2019:Content}, and Al-maneea\etal~\cite{Al-maneea:2019:Towards} optimized the layout arrangement of multiple-view visualization in infographics, magazines, visual analytics systems, respectively. 

This work focuses on performance dashboards, which are primarily used within industrial settings to present operational performance and hold a prominent position in the business intelligence market.
Despite their importance, research specifically targeting performance dashboards is limited.
To fill the gap, we introduce a performance dashboard design space that captures their common design patterns.

\subsection{Dashboard Generation and LLM-Powered Agents}
To lower the barrier to authoring dashboards, an increasing amount of work has focused on automating this process.
The application of rule-based methods integrates visualization knowledge into the design of metrics and the consequent chart recommendations~\cite{Wang:2020:DataShot,Zhao:2023:ChartStory,Shi:2021:Calliope,10.1145/2702123.2702149}. 
Researchers have explored machine learning-based methods for dashboard generation. 
Bidirectional long short-term memory models were applied in MultiVision~\cite{Wu:2022:MultiVision} for multiple-view visualization generation by combining single charts with high rank scores. 
DashBot~\cite{Deng:2023:DashBot} employed reinforcement learning methods integrating domain knowledge to dashboard generations. 
Additionally, natural language processing models can be applied to data feature extraction or recommendation next-step queries for dashboard exploration~\cite{Srinivasan:2019:Augmenting, wang:2022:interactive}. 
Data Illustrator~\cite{Liu:2018:DataIllustrator} proposed a vector drawing method that employs lazy data binding to visualization generation. 
LADV~\cite{Ma:2021:LADV} enables the identification and generation of dashboard templates from users' hand sketches, thus facilitating efficient ideation and authoring.
We continue to explore LLMs for the automation of dashboard authoring, focusing specifically on design prototyping and investigating dashboard generation via natural language input.

The powerful generative capabilities of LLMs have been successfully applied as intelligent question-answering systems in many domains, such as finance~\cite{wu:2023:bloomberggpt}, health~\cite{li:2023:does}, biology~\cite{zheng:2023:structureinformed}, and law~\cite{sun:2023:short}.
LLM-powered agents~\cite{10.1145/3586183.3606763,lin2025carbonsiliconcoexistcompete,10.1145/3526113.3545616} have demonstrated significant human-like decision-making capabilities and supported interaction with humans using natural language~\cite{zhou2024webarenarealisticwebenvironment,zhang2024simulatingclassroomeducationllmempowered,10.1145/3711066}.
Recently, researchers in the visualization community have also begun to integrate LLM-powered agents into the visualization creation process, leveraging their innovative and multimodal generative capacity~\cite{YE202443,10891192}.
Li\etal~\cite{li2024visualizationgenerationlargelanguage} evaluated the performance of LLMs in data visualization generation with Vega-Lite.
Prompt4Vis~\cite{li2024prompt4vispromptinglargelanguage} utilized an in-context learning method to push the boundaries of LLMs for text-to-vis tasks.
Compared to traditional machine learning or deep learning methods, LLMs' capability to generate coherent and contextual content~\cite{NEURIPS2022_9d560961} can facilitate reasoning in generative visualization, improving the interpretability of the design process.
LLM4Vis~\cite{wang2023llm4visexplainablevisualizationrecommendation} can recommend visualizations with human-like explanations.
ChartGPT~\cite{10443572} developed a step-by-step reasoning workflow to generate visualizations that answer users' data analytical problems.
Furthermore, LLM-powered agents also enable efficient and creative stylization by supporting text-to-style adaptation, automated aesthetic refinement, and personalized visual transformation.
LIDA~\cite{dibia2023lidatoolautomaticgeneration} utilized LLMs and IGMs to generate stylized and data-faithful infographics from datasets.
Xiao\etal~\cite{10296520} and Wu\etal~\cite{wu2023viz2vizpromptdrivenstylizedvisualization} embedded semantic context into charts based on generative AI.
Following this technical trend, we present a novel exploration within the text-to-vis domain and leverage the strong generation capabilities of LLMs to support an end-to-end dashboard generation workflow. 
We introduce a multi-agent pipeline that interprets natural language input and automatically completes chart type, layout, and styling, enabling efficient and accurate generation without manual authoring.

\subsection{Human-AI Collaboration for Design}
Human-AI collaboration for design has become a significant research frontier in HCI~\cite{10.1145/3290605.3300233}, ranging from AI-as-tool to genuinely co-creative, mixed-initiative systems~\cite{10.1145/302979.303030}.
AI can play a significant role in supporting designers throughout the design process by identifying contextual needs, externalizing tacit ideas, and exploring a broad design space through automatic variation~\cite{10.1145/3610217}.
FashionQ~\cite{FashionQ} harnesses the capabilities of AI to enhance creativity and support decision-making processes in professional fashion design.
ColorCook~\cite{colorcook} assists designers interactively in creating expressive and effective dashboard color schemes aligned with data semantics.
Vinci~\cite{vinci} takes a user-provided product image and taglines as input to automatically select suitable design elements and layout templates, producing an aesthetically coherent poster.
Designers engage not only as recipients of AI-generated outputs, but also as co-creators, who interact with AI systems to iteratively shape and refine design outcomes, rather than passively consuming results~\cite{10.1145/3613904.3642812,10.1145/3613904.3642726}.
AIdeation~\cite{Aideation} proposed a human-centered AI system that enables concept artists to efficiently explore creative directions through a flexible and iterative workflow.
Keyframer~\cite{tseng2025keyframerempoweringanimationdesign} introduces a natural language interface for motion design, aiming to make the animation process more intuitive and accessible. 
By enabling a dynamic feedback loop between user input and system response, it empowers animators to explore, iterate, and refine motion ideas with greater creative autonomy.
Uusitalo\etal~\cite{10.1145/3643834.3661624} indicated that industrial designers varied in both design tenure and prior exposure to GAI technologies, allowing for a nuanced understanding of how such tools are being integrated into real-world workflows.

The emergence of LLMs has further advanced human-AI collaboration in design. 
The powerful linguistic reasoning capabilities enable designers to articulate nuanced design intents through natural language, significantly lowering the interaction barrier in design workflows~\cite{10.1145/3581641.3584078,Wang_2024,Aideation}.
Chiou\etal~\cite{10.1145/3563657.3596001} indicated that integrating human and AI contributions within the design process facilitates new forms of self-expression and enriched modes of communication.
By combining image exemplars with textual cues, IntentTuner~\cite{IntentTuner} integrates user intentions into the fine-tuning workflow of text-to-image generative models, supporting intent-driven data augmentation.
While recent advances in LLMs have demonstrated strong potential in supporting creative tasks and accelerating design workflows, few tools have directly addressed the real-world demands of performance dashboard prototype design, often under tight timelines and evolving client requirements. 
To bridge this gap, we develop an interactive system, DashChat, that enables users to author performance dashboard prototypes through conversational interaction and facilitates an iterative, feedback-driven trial-and-error design process.

%% file: sections/3overview.tex
\section{Formative Study and Design Requirements}
\label{sec:overview}
In this section, we present a formative study that underpins our research. This process revealed a set of critical challenges in current practice. We derive six design requirements (R1-R6) directly from these findings, which serve as the foundational principles for our system's design.

\subsection{Methodology}
\label{sec:formativeStudy-methodology}
Our study centered on semi-structured interviews with 10 industry professionals to investigate the performance dashboard development lifecycle.

\subsubsection{Participants}
We recruited 10 participants~(6 males and 4 females; ages 29 to 41) through social media: 3 visualization researchers (VR1-VR3), 2 business analysts (BA1-BA2), 3 data analysts (DA1-DA3), and 2 visualization engineers (VE1-VE2). All participants had 5 to 10 years~(mean = 7.7,
SD = 1.64) of professional experience developing performance dashboards for enterprise and government clients.
\subsubsection{Procedure}
We conducted semi-structured interviews with each participant, lasting 1 to 1.5 hours. The interviews focused on workflow, prototyping practices, tool-related pain points, and communication challenges, which followed a standardized protocol focusing on:
(1) How stakeholders coordinate during the preparation phase of dashboard development.
(2) Current practices for rapid prototyping without access to real datasets.
(3) Pain points in existing tools when handling iterative design changes.
(4) How domain-specific requirements influence design decisions.
(5) Communication challenges between designers and engineers during data integration.
After the interviews, the insights were cross-validated with our firsthand experience collaborating with the DataV~\cite{Datav:2024:builder} team and by analyzing documentation from 3 of their recent projects.

\subsection{Findings}
Our analysis of the interview data revealed a significant gap between idealized academic pipelines and the realities of performance practice. We synthesized the findings into three core challenges that form the primary motivation.

\subsubsection{Workflow in Industrial Practice}
We first summarize this typical workflow. The process often begins with a \textbf{pre-data phase}, where business analysts and designers craft initial prototypes from vague client requirements, primarily for negotiation and securing contracts. This is followed by a highly \textbf{iterative design phase}, where these prototypes undergo numerous revisions based on stakeholder feedback, often still in the absence of real data. Finally, the workflow enters a \textbf{data integration phase}, where engineers connect the finalized designs to actual production data and orchestrate the interactive links among them.


\subsubsection{F1: The ``Pre-Data'' Prototyping Dilemma}
A dominant theme was the critical need to create prototypes before any real data is available, a phase driven by client negotiation.
\begin{itemize}
    \item \textbf{Client-Driven Need:} \textbf{BA1} highlighted the commercial pressure: ``\textit{Clients expect to see multiple design options before signing contracts, but we do not have access to their actual data yet.}''
    \item \textbf{Text-to-Visual Gap:} This phase begins with abstract, textual requirements. \textbf{BA3} noted the inefficiency of current tools: ``\textit{We need to create convincing demos based only on textual requirements. However, existing tools cannot translate this into visual components efficiently.}''
\end{itemize}

\subsubsection{F2: The Burden of Iterative and Non-Linear Design}
Participants consistently described the design process as chaotic and non-linear, characterized by constant iteration and unforeseen delays.
\begin{itemize}
    \item \textbf{Frequent Requirement Changes:} \textbf{DA2} shared a common experience: ``\textit{After initial prototyping with simulated data, clients frequently change requirements. Last month, we had to redesign 40\% of a manufacturing dashboard after three rounds of feedback.}''
    \item \textbf{Data Delays and Versioning Chaos:} \textbf{VE1} emphasized that even after a project starts, ``\textit{data delivery delays are common.}'' This, combined with frequent changes, leads to version control issues. ``\textit{Keeping track of different iterations is chaotic,}'' recalled \textbf{VE2}.
\end{itemize}

\subsubsection{F3: The Complexity of Domain-Specific Design}
A one-size-fits-all approach to design is insufficient. Prototypes must be convincing and adhere to specific domain conventions.
\begin{itemize}
    \item \textbf{Aesthetic and Layout Rules:} Participants noted that different domains have distinct visual languages. For instance, \textbf{DA1} mentioned that government dashboards require ``\textit{dark blue palettes with clear typography,}'' while VR2 stated that manufacturing dashboards need ``\textit{precise scale matching with live sensor data.}''
    \item \textbf{Need for Plausible Data Simulation:} To make prototypes believable, the simulated data must be meaningful. \textbf{DA1} explained, ``\textit{When creating prototypes, we often have to guess data distributions. We might simulate 10 different traffic flow patterns for a smart city dashboard just to show possible visualizations.}''
\end{itemize}

\subsection{Design Requirements}
\label{sec:requirment}
Grounded in the core findings identified in our formative study (F1-F3) and literature~\cite{Chen:2021:Composition, Bach:2023:Dashboard,li2024visualizationgenerationlargelanguage,10443572,wu2023viz2vizpromptdrivenstylizedvisualization,Ma:2021:LADV}, we synthesized our findings into six foundational design requirements (R1-R6):

\begin{enumerate}[label=\textbf{R{\arabic*}}, nolistsep]
    \item
    \textbf{Efficient prototyping with textual input.} To directly address the ``pre-data'' dilemma and the text-to-visual gap \textbf{(F1)}, the system must offer a direct text-to-prototype methodology. This enables users to translate textual ideas into tangible formats without preparing data, thus streamlining the labor-intensive initial ideation phase.
    \item
    \textbf{Visual mapping with simulated domain-relevant data.} To create convincing mock data in the absence of real data \textbf{(F1)} and to meet the need for plausible and domain-specific visualizations \textbf{(F3)}, the system must populate charts with meaningful simulated data. The data should reflect the essentials of the user's input and the relevant domain context, making the prototype a more faithful representation of the final product.
    \item
    \textbf{Strategic layout planning.} In line with the need for strategic designs \textbf{(F3)}, the system should thoughtfully arrange components based on their interrelationships and established layout patterns to ensure visual clarity and effective data communication.
    \item
    \textbf{Aesthetic and stylistic designs.}  To adhere to the strict domain-specific conventions \textbf{(F3)}, the system must intelligently apply aesthetic elements. This includes selecting appropriate color patterns and incorporating common embellishments to ensure the prototype is not only functional but also stylistically and aesthetically coherent.
    \item
    \textbf{Optimization of LLMs generation performance.} The inherent complexity of fulfilling the above requirements (R1-R4) demands a robust technical foundation. Therefore, the system must employ strategies to optimize the performance of LLMs, supplying it with the necessary domain knowledge to ensure both efficient and high-quality generation.
    \item
    \textbf{Augmenting interface interactivity}.  To combat the burdensome iterative workflow and versioning chaos \textbf{(F2)}, the interface must support fluid and continuous refinement. This involves enabling multi-turn conversational input for complex changes while also offering direct manipulation controls for quick, targeted adjustments, thereby enhancing user efficiency and control.
     
\end{enumerate}


%% file: sections/3.5design_pattern.tex
\section{Survey on Performance Dashboard Design}
\label{sec:design-pattern}
To discover the common design practices in practical performance dashboards, we conduct a formative study based on a collection of examples from various real industrial scenarios. 

\subsection{Methodology}
\begin{figure}[ht]
\centering
\includegraphics[width=\linewidth]{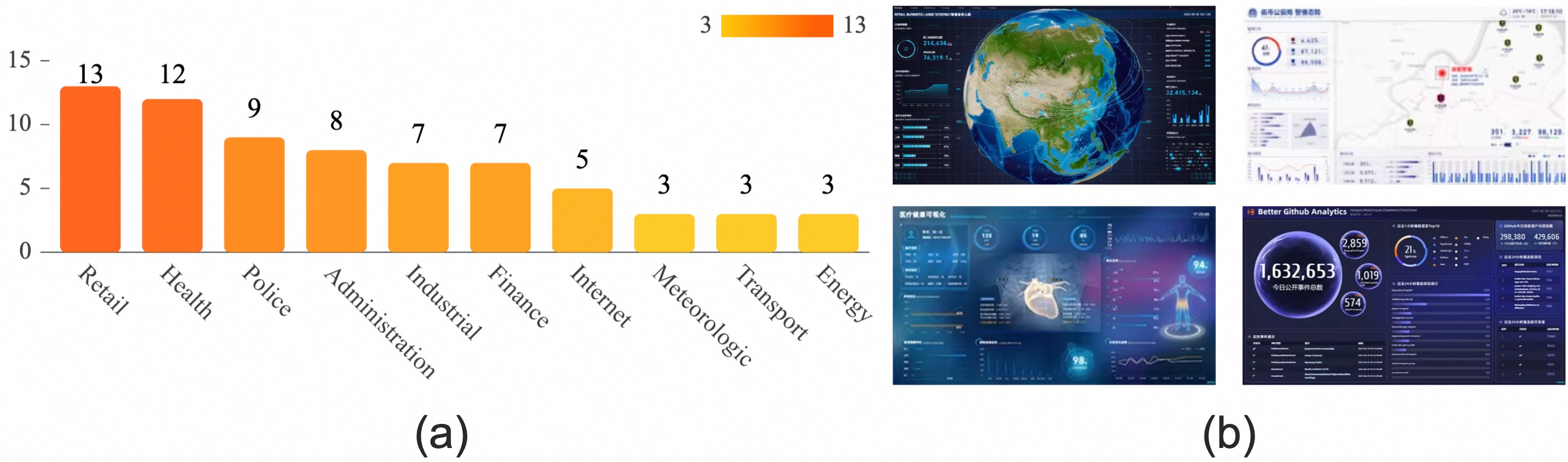}
\caption{(a) Distributions of the domains. (b) The sample dashboards with different domains: Retail, Health, Police and Administration.}
\label{fig:domainDistribution}
\end{figure}

Although numerous dashboard sources are available online, we focus on the specific presentation of industry dashboards. 
Dashboard examples are selected from industry-based templates constructed by DataV~\cite{Datav:2024:builder} and other commercial dashboard authoring tools. 
$146$ performance dashboards created by professional designers, combining aesthetics and effectiveness, are first collected. 
Then, those focusing solely on display rather than data presentation, and those with only a single view are excluded. 
We finalize a dataset with $114$ samples, covering a wide variety of domains, shown in Fig.~\Ref{fig:domainDistribution}~(\eg retail~(11.4\%; 13), health~(10.5\%; 12), police~(7.9\%; 9)).

As our summary of common design practices inform LLM-powered agents in the system pipeline~(\textbf{R5}), it complements existing dashboard design patterns rather than providing a complete taxonomy~\cite{Card:1999:Readings, Chen:2021:Composition, Bach:2023:Dashboard}.
Bach\etal~\cite{Bach:2023:Dashboard} argued that dashboard content components follow common structures, categorized into page layout, screenspace, content structure, interactions, and color scheme.
We conduct discussions with the team members from DataV~\cite{Datav:2024:builder} and learn that performance dashboard design patterns aim to maximize screen use for visuals while minimizing audience interpretation bias.
Meanwhile, most clients attempted to apply a dashboard for data insight presentations. 
Thus, we only consider the display relations of views, excluding their interaction relations. 

A qualitative analysis of the dataset is conducted for extracting views and dashboard designs from the aspects including \textbf{View Type}, \textbf{View Layout}, \textbf{Analysis Task}, and \textbf{Embellishment}. These can be separated into dashboard and single-view designs.
Inspired by Chen\etal~\cite{Chen:2021:Composition}, we first examine the relationships between all the views in dashboards, including their types and the position distribution of these views, namely the view layout. Then, we delved into every element in each view to determine the relationship between the analysis task of the view and the typical display form. Additionally, we investigate the application of embellishments in single views.


\subsection{Dashboard Design}
We explore the structure and layout of performance dashboards, drawing on the concepts of \textbf{View Type} and \textbf{View Layout} from Chen \etal~\cite{Chen:2021:Composition}. These attributes define the chart types and spatial organization of views, indicating the content relationships and layout hierarchies within dashboards.

\subsubsection{View Type}
\begin{figure}[ht]
\centering
\includegraphics[width=\linewidth]{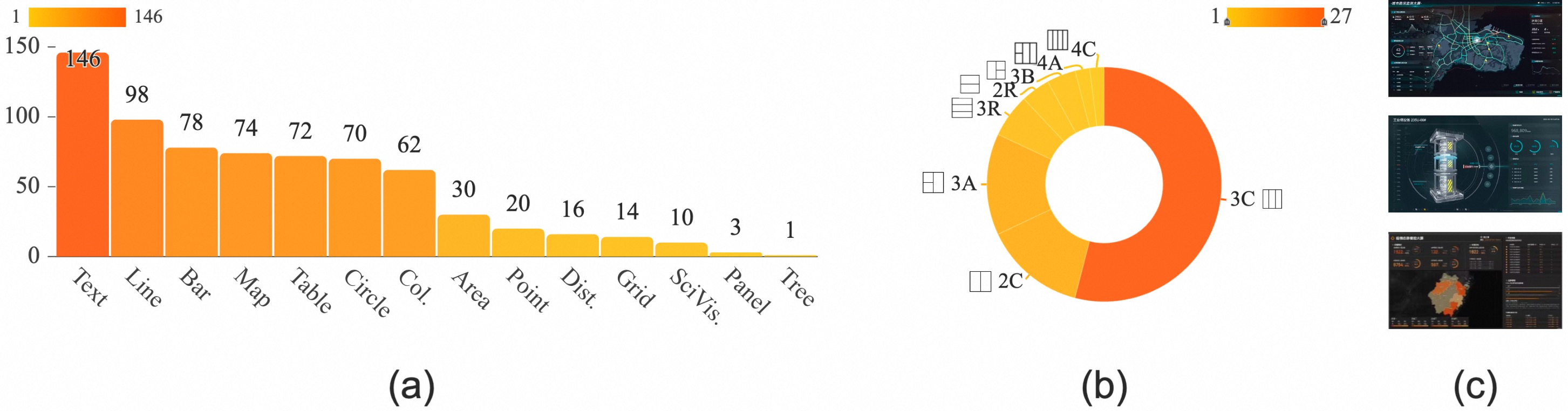}
\caption{(a) Distributions of view types.~(b) Distribution of dashboard layout in 2-Level hierarchy.~(c) The sample dashboards within different layout configurations: 3C, 2C, 3A.}
\label{fig:typeDistribution}
\end{figure}

View Type refers to the visual mapping method used to transform data into a visual form. We base our selection of supported chart types on the comprehensive taxonomy proposed by Chen\etal~\cite{Chen:2021:Composition}, which identifies 14 common types for multi-view visualizations.
To focus our generative model on the most relevant and prevalent chart types for performance dashboards, we first conducted a preliminary analysis of a corpus of existing dashboard examples. Our survey revealed that certain types, such as Panel (often used for interactive controls) and Tree (for hierarchical data), appear with extremely low frequency in typical dashboard contexts (see Fig.~\ref{fig:typeDistribution}(a) for a visual summary of this distribution). This empirical finding, combined with the limited utility of interactive controls like Panel in our static prototype generation context, led us to exclude both from our scope.
Additionally, while SciVis (Scientific Visualization) is present in some specialized dashboards, its highly domain-specific and custom nature makes it unsuitable for generalized automated generation at this stage. Therefore, while our analysis acknowledges the full spectrum of chart types, our prototype generation process concentrates on the remaining 11 types that are most pertinent to performance dashboards, such as bar charts, line charts, and maps.

\subsubsection{View Layout}
View Layout is defined by how views are positioned and grouped within a dashboard~(Fig.\ref{fig:typeDistribution}(b)). Building on the concept of \textit{bounding box}'', which is the position and size of a view in display space, we formalize layouts based on bounding box arrangements~\cite{Chen:2021:Composition}. Following Chen \etal's hierarchical approach, we adopt a two-level hierarchy where level-2 nodes often group views of the same type, traditionally termed as ``small multiples'' by Tufte~\cite{Tufte:1986:The}. 
However, we hold an alternative opinion that the views of the same visual type can present a specific analysis presentation for a particular analytical perspective that cannot be simply concluded as the level-2 nodes in the dashboard structure. Meanwhile, there also exist multiple neighboring visual types for conveying the same data insights. Therefore, we extend the definition of `` small multiple''  that it qualifies as long as it conveys a unified data insight, even if the charts differ in type. This broader interpretation enhances visual communication for specific analytical tasks, as detailed in~\cref{sec:AnalysisTask}.

To model dashboard layouts, we retain the hierarchical structure but refine the computation of layout relationships. A node representing multiple views can split into child nodes when the splitting direction changes; otherwise, sibling nodes represent individual views. Our analysis reveals that 80\% layout trees derived from samples exhibit a two-level hierarchy (Fig.~\ref{fig:typeDistribution}(b)), reflecting a preference for concise designs that maximize readability. Based on this insight, we incorporate four primary layout patterns into our generation process.

\subsection{Single View Design}
We disaggregate each view of the dashboard into individuals to investigate the features of the single-view display.
We extract $674$ individual views from $114$ dashboard examples. Since our goal is to investigate the views which contain multiple visual elements, we extend the definition of \textit{small multiples} that are considered as a single view which is made up of multiple visual elements that correspond to the same data insights or analysis tasks. 
Therefore, $470$ views are removed as they only contain single visualizations in views, which do not fulfill the definition of \textit{small multiples}. Therefore, we hold $204$ ``small multiples'' as the sample data.

\begin{figure*}[ht]
\centering
\includegraphics[width=\linewidth]{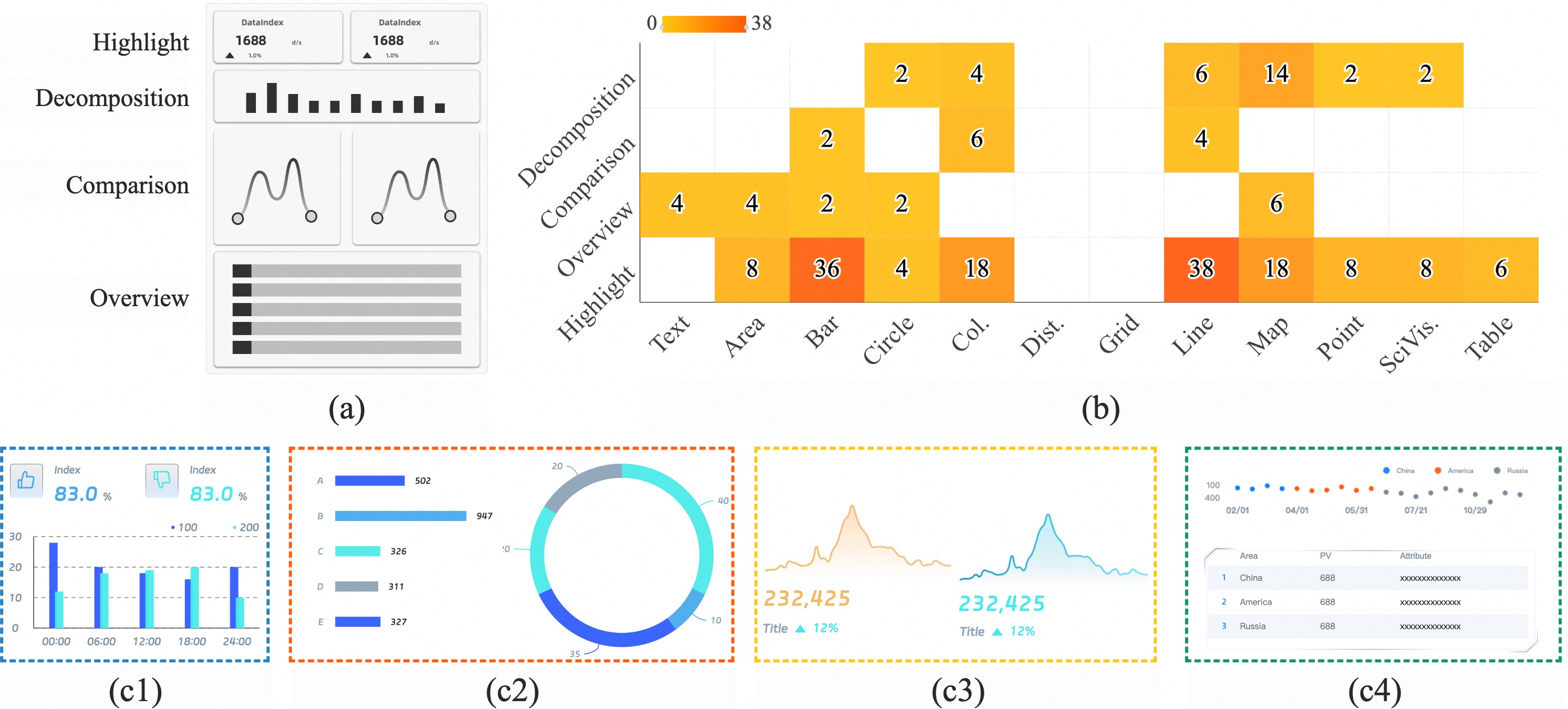}
\caption{(a) Layout Paradigms for Small Multiples in Different Analysis Tasks.~(b) Distribution of Different Analysis Tasks in Dashboards.~(c) The sample small multiples within different Analysis Tasks: (c1) Highlight, (c2) Decomposition, (c3) Comparison, and (c4) Overview.}
\label{fig:smDistribution}
\end{figure*}

\subsubsection{Analysis Task}
\label{sec:AnalysisTask}

We believe that individual views with multiple visualizations simultaneously serve specific analysis tasks (Fig.\ref{fig:smDistribution}(a)) while communicating unified data insights. We categorize these views based on the analysis task criteria from DataShot~\cite{Wang:2020:DataShot}. However, survey results from DataShot suggest no significant correlation between visualization types and low-level analysis tasks~\cite{Amar:2005:Low}, as designers often innovate in chart selection. Instead, plotting data characteristics are more closely tied to visualization types, so we exclude visualization types from the consideration. Based on the survey findings, we summarize four primary categories of analysis tasks:

\textbf{Highlight} emphasizes specific values, extremes, proportions, or outliers within a dataset under defined criteria~\cite{Wang:2020:DataShot}. For instance, displaying ``the number of flights departing on a given day'' in aviation.

\textbf{Decomposition} can be conceived as the tasks that combine two or more data attribute sets, trying to explore the relationship among them, also defining as \textit{correlate}. From the examples, this mostly appears when one of the sets is a subset of the other. For instance, this can be applied when ``the different profit shares of a given company based on displaying the profitability of all companies'' in the field of finance.

\textbf{Comparison} emphasize differences between multiple data attributes sharing similar features or significance, or differences across target objects for the same attribute. For instance, this occurs when analyzing ``the number of daytime and nighttime vehicle trips'' in transportation.

\textbf{Overview} provides a comprehensive view of datasets covering multiple attributes or complete plotted data, prioritizing completeness and accuracy. Contrary to \textbf{Highlight} task, this method places greater emphasis on data completeness and accuracy. For example, presenting ``prizes won by sports teams and their distributions'' in sports.

\subsubsection{Embellishment}
During the disaggregation process, we identified a significant number of embellishments such as borders, dividers, and icons in the dashboard examples, totaling $1028$ instances. These elements play a key role in enhancing visual association and improving topic comprehension~\cite{chen:2022:vizbelle}. Although embellishments are not formally defined in data visualization, they are generally regarded as non-essential graphical components that do not represent data directly but serve aesthetic or metaphorical purposes~\cite{bateman:2010:useful}. To analyze their usage, we categorized embellishments in $503$ single view examples based on their types. Due to their similar functional roles, borders and dividers were grouped as \textbf{Borders}, while icons were analyzed separately. The dimension includes two primary elements: \textbf{Borders} and \textbf{Icons}.

\textbf{Borders} provide recognizable margins to include all visual elements in views. Apart from aesthetic additions, the elements also indicate the relevance of the charts included and visually guide the audience to see the entire view as a whole. 

\textbf{Icons} focus on providing contextual information of the plotting data that help communicate abstract information into concrete mainly for topic comprehension enhancement.

%% file: sections/4system.tex
\begin{figure*}[ht]
\centering
\includegraphics[width=\textwidth]{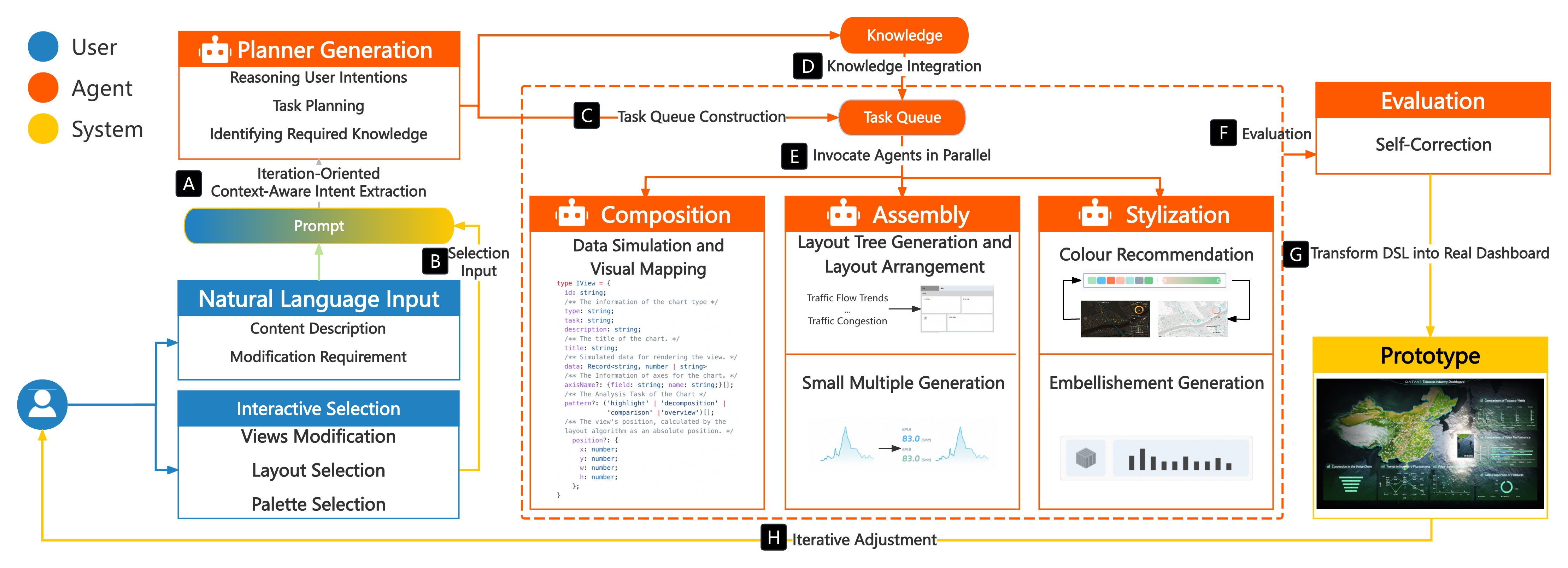}
\caption{System Pipeline.~(A) Textual Input to extract tasks.~(B) Selection input to directly assign tasks.~(C) Planning tasks to determine the task workflow.~(D) Integration of specific domain knowledge to augment LLM agent.~(E) Invocating Agents in Parallel.~(F) Self-evaluation the generation result.~(G) Transform generated output into a real prototype.~(H) Adjust the prototype iteratively.}
\label{fig:pipeline}
\end{figure*}

\section{System Pipeline}
\label{sec:pipeline}
The DashChat system is designed as a multi-agent pipeline that translates conversational user requests into performance dashboard prototypes (see Fig.~\ref{fig:pipeline}). The pipeline generates an initial prototype as a reference for iterations and a visual medium for idea exchange, rather than a final product. Drawing insights from the survey in~\cref{sec:design-pattern}, we summarize key features of the performance dashboard derived from real industrial scenarios. 
The pipeline is implemented with Node.js, backed by Qwen2.5-Max~\cite{qwen25}.
After piloting several commercial and open-source LLMs, we chose it for its strong instruction-following capability, competitive latency, and favorable unit cost.
We present the prompt templates and some technique details in the supplementary materials.

\subsection{User Input and Task Creation}
This module forms the foundation of our pipeline, translating inputs into actionable tasks for the LLM-powered agent to generate prototypes (\textbf{R1}). It processes both natural language and interactive selections to formalize user intent.

The flexibility of natural language often results in ambiguous, context-dependent commands (\eg ``\textit{change that chart}''), turning intent parsing into a high-risk inference problem. To address this, we introduce \textbf{Iteration-Oriented Context-Aware Intent Extraction} (Fig.~\ref{fig:pipeline}(A)), which reframes parsing as a constrained and verifiable generation task. Our approach utilizes a structured prompt that elicits the Chain-of-Thought~(CoT) strategy~\cite{touvron2023llama}. Our prompt integrates conversational history, a JSON-style domain-specific languages~(DSL)~(see supplementary materials) of the current dashboard state and the target visualization elements~(\eg element ids), and several few-shot examples. These examples demonstrate not only the target JSON format but also the desired step-by-step reasoning process~\cite{Wei2022Chain-of-thought}. 
It ensures a reliable and predictable foundation for the entire workflow. The module also supports \textbf{Selection Input} (Fig.~\ref{fig:pipeline}(B)). It allows users to select predefined templates or widgets, bypassing language ambiguity for rapid and error-free prototype generation.

\subsection{Task Planning and Knowledge Integration}
\label{sect:pipeline-Task-Planning}

The Task Planning Module organizes and prioritizes tasks generated by the Task Creation Module, ensuring logical and efficient execution. This module addresses two critical difficulties that not only ensure a logical execution order for complex requests, but also infuse the generation process with specialized domain knowledge. We leverage prior experience in dashboard construction to manage task dependencies, isolate independent tasks, and prioritize interdependent ones. Additionally, it integrates Retrieval-Augmented Generation~(RAG)~\cite{2023arXiv231210997G} by retrieving contextual knowledge from our knowledge base to support task execution. We utilize a text embedding model to embed the dataset mentioned in \cref{sec:design-pattern}, as the knowledge documents, into vectors, enabling efficient retrieval by the LLM agent.

\begin{figure}[ht]
\centering
\includegraphics[width=\linewidth]{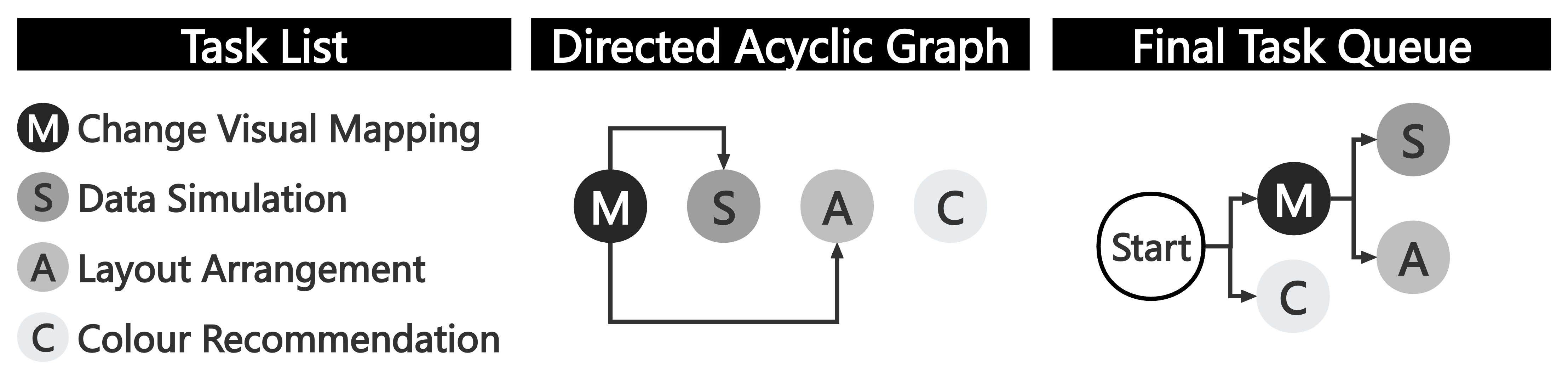}
\caption{An Example of Task Dependency Analysis and Sequencing. After obtaining the initial task list [M, S, A, C], the tool analysis reveals that tasks A and S are dependent on the output of task M, while task C is an independent task. Then, a DAG is constructed to represent the task dependencies, and the final task execution queue is determined. Tasks M and C are executed in parallel; After completion of M, tasks S and A are executed synchronously.}
\label{fig:pipeline-DAG}
\end{figure}

\textbf{Task Dependency Analysis and Sequencing}: 
A single user request often decomposes into multiple sub-tasks with inherent dependencies (\eg a chart must be created before its style can be modified). Executing these tasks in an arbitrary order can lead to logical errors and failed generations. To resolve this, the module first performs a task dependency analysis, modeling the relationships between tasks as a Directed Acyclic Graph (DAG) that defines a fixed execution order (Fig.~\ref{fig:pipeline-DAG}). Actions corresponding to each task, defined as operations necessary for task completion, are added to an action group in sequence, ready for execution. This guarantees a robust and logical workflow, preventing cascading failures and ensuring that each step builds upon a valid prior state. Subsequently, the \textbf{Task Queue Construction} process (Fig.~\ref{fig:pipeline}(C)) traverses this DAG to generate a linear, ordered execution queue. This ensures that all prerequisite tasks are completed before their dependent tasks begin, transforming an unordered task set into a structured plan and guaranteeing a robust workflow that prevents cascading failures.

\textbf{Knowledge Integration} (Fig.~\ref{fig:pipeline}(D)): To compensate for the lack of the specialized knowledge of data visualization best practices from generic LLMs, we employ Retrieval-Augmented Generation (RAG) strategy (Fig.~\ref{fig:pipeline}(D)). We constructed a curated knowledge base from the dataset in \cref{sec:design-pattern}, containing structured design examples in JSON format. Each example details a description, recommended charts and their analysis task, and style choice. Based on the current task, the agent formulates a semantic query and performs a vector search to retrieve the most relevant ones. The top three retrieved examples are then prepended to the LLM context. This approach grounds the creativity of LLMs in a corpus of proven best practices, ensuring the resulting dashboard prototypes are not only plausible but also professionally sound and aligned with industry standards.

\subsection{Task Implementation}
The module is a critical component of our pipeline, executing the ordered workflow generated by the Task Planning Module, ensuring efficient task completion. It processes independent tasks in parallel while respecting dependencies and invoking dependent tasks asynchronously (Fig.~\ref{fig:pipeline}(E)). To support this, the module integrates specialized sub-agents for prototype generation: \textbf{Composition}, \textbf{Assembly} and \textbf{Stylization}. The tools use dedicated memory to store and pass intermediate results to dependent tasks, ensuring seamless workflow integration.

\subsubsection{Composition}
The Composition Agent is responsible for the generation of individual data visualization components such as charts and KPIs, fulfilling the requirement (\textbf{R2}). Its primary function is to transform a single high-level analytical goal from the task queue into one or more meaningful and structured chart representations based on our DSL. 


 \textbf{Data Simulation and Visual Mapping}: Generating plausible data is critical for a convincing prototype. Our process mitigates hallucination by separating schema definition from data generation:

 \begin{itemize}
    \item \textbf{Schema Definition via LLM:} Based on the display task, the LLM is prompted to define a data schema in a structured JSON format. For instance, there is a task as ``show product category distribution''. This schema specifies column names (\eg product name, sales volume), data types (\eg String, Integer), and statistical properties (\eg a value range for sales volume, a list of possible product name categories). To guide the LLM, the prompt includes few-shot examples and leverages insights from datasets like NvBench~\cite{luo:2021:Synthesizing} to inform realistic data distributions.
    \item \textbf{Deterministic Data Generation:} With the schema defined, a separate, deterministic script generates the data points according to the specified constraints. This two-step process prevents the LLM from directly fabricating numerical values, significantly reducing the risk of nonsensical or random data. The LLM's role is confined to creative schema design, while a predictable process handles the generation. Finally, the tool maps the generated data fields to visual encodings (\eg product name to the x-axis, sales volume to the y-axis).
\end{itemize}
 
 The output of the Composition Agent is therefore a collection of structured DSL snippets, each representing a fully specified chart ready for the subsequent Assembly stage.

\subsubsection{Assembly}
The agent organizes visual elements systematically to ensure clarity, readability, and alignment with user requirements (\textbf{R3}). It integrates \textbf{Layout Tree Generation and Arrangement} and \textbf{Small Multiple Generation}, guided by prior research and empirical data (\cref{sec:design-pattern}).

\textbf{Layout Tree Generation}: The tool constructs a hierarchical layout tree to define spatial relationships between views. Inspired by composition pattern analysis~\cite{Chen:2021:Composition}, the layout tree is modeled as a multilevel hierarchy. 
The structure is as follows:
\begin{equation}
Layout: = \{\underbrace{view_{1}, [view_{2}, ...,view_{i}], view_{n} }_{\leq 4}\}
\end{equation}
where the two-tuple \(view_{i}:= (type_{i}, [width_{i}, height_{i}])\), symbolizes the main chart type applied in the \(view_{i}\) and the relative width and height of the view in the display screen.
Particular, the results of the survey~(Fig.~\ref{fig:typeDistribution}(b)) indicate that the layout tree gained from the samples is a 2-level hierarchy where the level-1 node quantity mainly ranges from 2 to 4, implying that the layout designs for industry dashboards are as concise as possible to ensure readability. Therefore, when calculating the layout tree for views, we only consider a tree with a 2-level hierarchy and at most 4 level-1 nodes.

When it comes to considering the pairwise relationships between two view types, we believe that the total number of views in a dashboard influences the information amount generated when the views are selected or laid out. A dashboard with a greater number of views means that the likelihood of a particular chart type being applied to it is more uncertain, which contains more information. Therefore, contrary to the application of conditional probability in the work~\cite{Chen:2021:Composition}, we apply the concept of information entropy and conditional entropy to represent the relationship:
\begin{multline}
H(type_{i} \mid type_{j}) = \\-\sum_{type_{i}, type_{j}}P(type_{i}, type_{j})log P(type_{i} \mid type_{j})
\end{multline}
where the value represents the entropy of \(type_{i}\) appears in dashboard when there exist \(view_{j}\) in the screen. Therefore, the lower \(Relation_{ij}\) implies the higher the correlation between \(type_{i}\) and \(view_{j}\) since it represents the lower the uncertainty in the emergence of the chart \(type_{i}\). We define the relationships between a view and node that may contain multiple views from the layout tree at level-1 as:
\begin{equation}
Relation_{ij}: = \min_{type_{k} \in node_{j}}H(type_{i} \mid type_{k})
\end{equation}
where \(type_{k}\) represents the various chart types included in \(node_{j}\). Based on the above two calculations, we propose the layout tree algorithm for layout construction for the following arrangement. We present the algorithmic pseudocode for constructing the layout tree in the supplementary materials.

\textbf{Layout Arrangement}: After constructing the layout tree, the tool arranges views in three steps:  The tool identifies the most representative view, which holds the largest proportional size, and models the relative position of each view on the screen \(S\) as \(S(view_{i}):=\{p_{i1}S_{1},...,p_{in}S_{n}\}\), where \(p_{ik}\) represents the proportion of overlapping space \(S_{k}\). By comparing positions, the module selects row-based or column-based layouts for level-1 nodes. View sizes are adjusted to reflect their importance, calculated by accumulating the values of the \(p_{i}\) set for all views within a node. The space of each view is allocated proportionally, ensuring size reflects significance. Finally, the tool constructs level-2 nodes by slicing the underlayer orthogonally to the level-1 split, highlighting the hierarchical relationships within the layout tree and reducing the cognitive load on users when interpreting the dashboard. View sizes within level-2 nodes are determined based on common dimensions and task-specific requirements, ensuring visual coherence.

\textbf{Small Multiple Generation}: To support complex analytical narratives, the tool supports small multiples, which are individual views containing multiple visualizations for specific analysis tasks. Based on our survey, we defined four common analytical patterns (Fig.~\ref{fig:smDistribution}(a)): (i) \textbf{Comparison} views use side-by-side charts with consistent types but distinct colors. (ii) \textbf{Highlight} views incorporate components like number flops or gauges to emphasize key metrics. (iii) \textbf{Overview} views employ tables or comprehensive charts to present a complete dataset. (iv) \textbf{Decomposition} views combine multiple chart types to facilitate hierarchical data exploration.

\subsubsection{Stylization}
The Stylization agent enhances the aesthetic appeal and functional clarity of dashboards using color patterns and embellishments. We apply LLMs for color recommendations and Image Generation Models~(IGMs) for embellishments, to ensure visually engaging and contextually consistent designs (\textbf{R4}).

\textbf{Color Recommendation}: The agent applies a global color palette, dynamically recommended by the LLM based on user input or thematic keywords. It distinguishes between \textbf{Categorical Palettes} for discrete data and \textbf{Sequential Palettes} for continuous data.

\textbf{Embellishment Generation}: The tool extracts a theme color, translates it into natural language prompts and uses IGMs~(Stable Diffusion v3~\cite{diffusion}) to generate design elements including borders and icons. To improve output quality, we fine-tuned the IGM on a custom dataset of dashboard components, creating specialized LoRa models for specific element types~(see supplementary materials).

\subsection{Result Evaluation and Iterative Adjustment}
This module serves as the final checkpoint in the pipeline, ensuring that the generated performance dashboard prototype meets the expectations and adheres to the quality standards. Here we integrate proactive constraints with post-hoc validation, addressing the difficulty of the creation of factually incorrect or nonsensical content from LLM.

\textbf{Result Evaluation}: As described in the Composition agent, the use of a strict DSL prevents the LLM from generating arbitrary structures or commands. In addition, by separating schema definition from data generation, we prevent the LLM from directly inventing numerical data, a primary source of hallucination. In this agent, before rendering the prototype, it applies automated evaluation on the abstract structure: (i) It validates the generated JSON structure against its schema, ensuring all required fields are present and correctly typed. (ii) The tool (Fig.~\ref{fig:pipeline}(F)) checks if the chosen chart type is appropriate for the characteristics of data (\eg if a line chart is used for non-temporal categorical data). If the evaluation identifies deficiencies, including structural errors or semantic inconsistencies, the pipeline automatically triggers a refinement loop by returning the task to the Planning module with specific feedback for correction.

\textbf{Prototype Construction} (Fig.~\ref{fig:pipeline}(G)): Once passing the evaluation, the abstract structure is translated into a real performance dashboard prototype. This tool bridges abstract structure and practical implementation, providing users with a realistic preview of the final product.

\textbf{Human-in-the-Loop Validation and Iterative Adjustment}: The pipeline is designed for iterative collaboration, not one-shot generation (Fig.~\ref{fig:pipeline}(H)). The generated prototype is a proposal for the designer to inspect. The designer, as a domain expert, is best equipped to identify subtle hallucinations, such as fabricated metrics that look plausible but are contextually meaningless, or inappropriate visual metaphors. The iterative adjustment process (\textbf{R6}) empowers the user to correct these issues conversationally, ensuring the final prototype is both structurally sound and semantically valid.

%% file: sections/4.5interface.tex
\begin{figure*}
  \centering
  \includegraphics[width=0.9\textwidth]{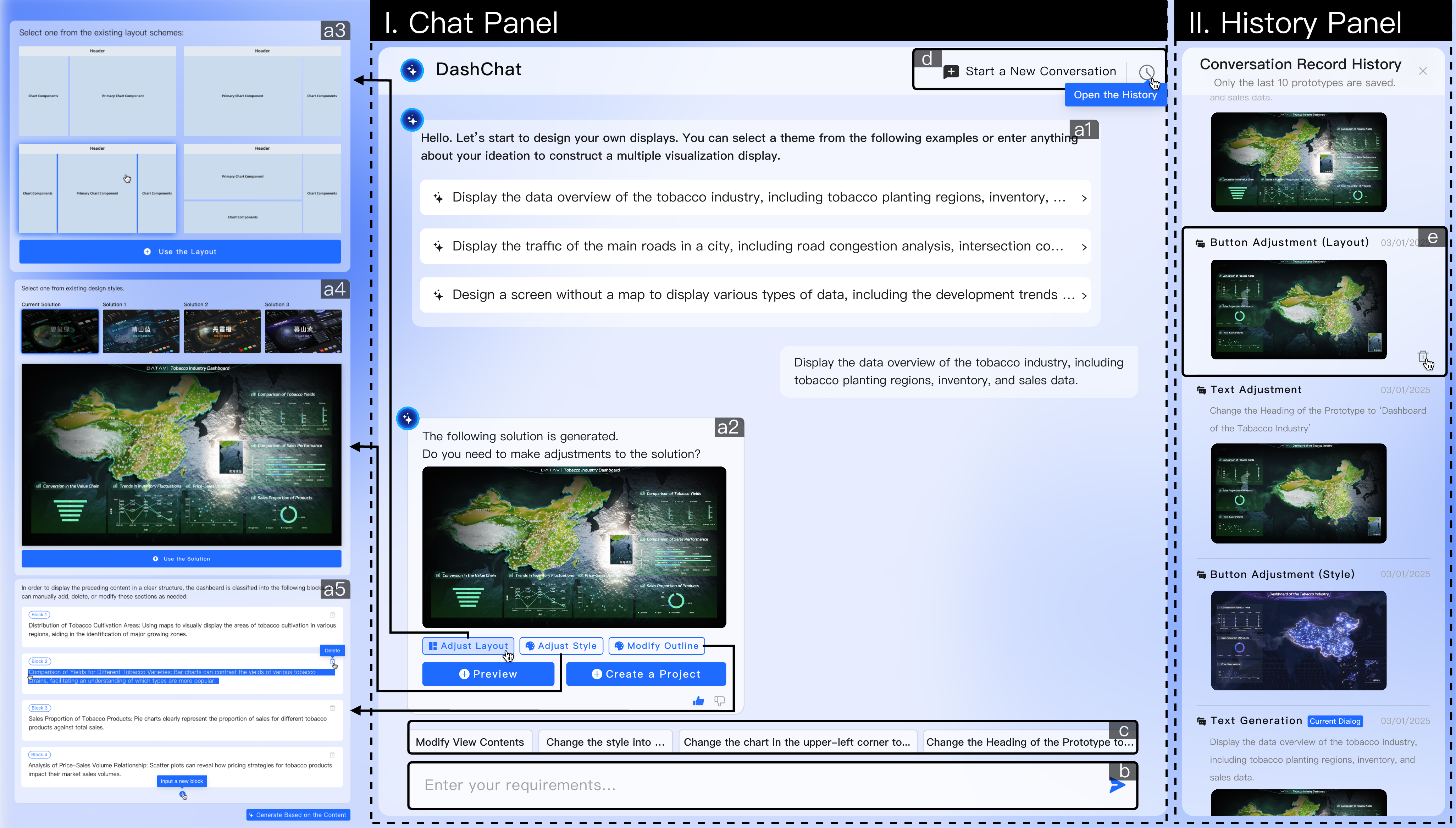}
  \caption{The system interface of DashChat. (I) The Chat Panel enhances communication between users and the system 
by enabling open-question answering and providing multiple potential selections in chat bubbles, allowing DashChat
users to clarify their needs through natural language textual input and intuitive interaction.
The functionalities of different bubbles are further elaborated in~\cref{sec:interface}.
 (II) The History Panel records the changes of the generated prototypes, where users can review their previous specifications and choose the prototype that best meets their satisfaction.
 }
  \label{fig:interface}
\end{figure*}

\section{Interface and Case Study}\label{sec:interface}
DashChat is designed to facilitate seamless interaction between users and the system, enabling efficient and intuitive generation and modification of prototypes. The interface consists of two primary panels: \textbf{Chat Panel} and \textbf{History Panel}. Each view is crafted to support a natural language-driven workflow, provide real-time feedback, and allow users to iteratively refine their designs~(\textbf{R6}).
The interface is written in TypeScript.
To illustrate the practical application and workflow of DashChat, this section presents a detailed case study following a designer, Nemo, as he undertakes the task of creating a high-fidelity prototype for a manufacturing monitoring dashboard. The scenario involves a tight deadline and evolving requirements, conditions under which the efficiency and iterative capabilities of DashChat are particularly valuable.

\subsection{Phase 1: Initial Prototype Generation}

Upon initiating a new project in DashChat, Nemo is presented with the Chat Panel. Faced with the initial ``blank canvas'' problem, he immediately benefits from the \textbf{Content Suggestion Bubbles}~(Fig.~\ref{fig:interface}(a1)) displayed above the chat input area. Instead of formulating a complex initial prompt, Nemo selects a predefined option that aligns with his project domain: ``\textit{Tobacco Industry Supply Chain}''.

This single click serves as the initial command. The system processes the request and, within seconds, generates a \textbf{Prototype Display Bubble} in the chat interface~(Fig.~\ref{fig:interface}(a2)). This bubble contains a foundational dashboard prototype, complete with relevant data visualization components such as Key Performance Indicator cards, a line chart for production trends, and a status overview for different assembly lines. This initial step effectively provides a structured starting point, allowing Nemo to bypass the time-consuming process of building a basic layout from scratch~(\textbf{R3}).

\subsection{Phase 2: Iterative Design Refinement}
With a baseline prototype established, Nemo proceeds with a series of iterative refinements using a combination of natural language commands and guided interactive elements~(\textbf{R1, R6}).

\subsubsection{Layout Modification:} The default two-column layout is not optimal for the intended display environment. Nemo seeks to change this. He clicks on the ``Modify Layout'' interactive button located within \textbf{Prototype Display Bubble}~(Fig.~\ref{fig:interface}(a2)). This action triggers the appearance of a \textbf{Layout Selection Bubble}~(Fig.~\ref{fig:interface}(a3)), which presents several predefined grid and column templates. Nemo selects a three-column layout. The system instantly re-renders the prototype, automatically and intelligently repositioning the existing chart views into the new layout structure, demonstrating the capability for rapid structural adaptation~(\textbf{R3}).

\subsubsection{Aesthetic Enhancement:} To align the prototype with the client's branding, which calls for a dark, high-tech aesthetic, Nemo interacts with the ``Modify Design Style'' button. This opens the \textbf{Design Style Recommendation Bubble}~(Fig.~\ref{fig:interface}(a4)). He is presented with a gallery of visual style previews, such as ``\textit{Serene Mountain Blue},'' ``\textit{Alpine Cloud White},'' and ``\textit{Gemstone Green}.'' After previewing the options, he selects ``\textit{Serene Mountain Blue}.'' The system applies the style globally, updating the background color, font styles, and chart color patterns across the entire prototype in real-time, showcasing an efficient method for exploring and applying visual themes~(\textbf{R4}).

\subsubsection{Content Specificity and Addition:} Nemo thinks that the prototype now requires more specific content. First, a ``\textit{Production Volume Pie Chart}'' needs to be changed to represent ``\textit{Defect Rate by Production Line}''. Nemo clicks the ``Modify Content'' button, which activates the \textbf{Content Modification Bubble}~(Fig.~\ref{fig:interface}(a5)). This bubble displays an interactive outline of all chart views in the prototype. Nemo locates the node corresponding to the pie chart and directly edits its title to ``\textit{Defect Rate by Production Line as a bar chart}''. The system interprets this change, replacing the pie chart with a bar chart populated with placeholder data relevant to defect rates. This demonstrates a structured and error-reducing method for content editing~(\textbf{R2}).

Second, Nemo need to add a new metric, ``\textit{Overall Equipment Effectiveness}''. For this, Nemo utilizes the \textbf{Chat Box}~(Fig.~\ref{fig:interface}(b)), typing the natural language command: ``\textit{Add a radar chart in the bottom right section showing Overall Equipment Effectiveness}''. The system parses this instruction, identifies the target location and chart type, and generates a new \textbf{Prototype Display Bubble} with the updated design, which now includes the requested radar chart. This free-form input highlights the conversational interaction capabilities of our system~(\textbf{R1, R6}).

\subsection{Phase 3: Managing Design History and Reversion}
During a review, a stakeholder expresses a preference for a layout that was explored earlier in the session. Instead of attempting to recall the exact previous configuration, Nemo utilizes the \textbf{Additional Interaction Options} in the top-right corner~(Fig.~\ref{fig:interface}(d)) to toggle the visibility of the \textbf{History Panel}.

The \textbf{History Panel} displays a chronological sequence of \textbf{Prototype Generation Record Cards}~(Fig.~\ref{fig:interface}(e)). Each card contains a thumbnail preview of the prototype at that stage, a timestamp, and the specific modification that was performed~(\eg ``\textit{Modified layout}'', ``\textit{Text-Based Generation}''). Nemo quickly scrolls through this visual timeline, identifies the card corresponding to the desired layout, and clicks on it. This action allows him to either revert to this previous state or use it as a new starting point for a different design branch. This feature provides a comprehensive audit trail and robust version control, critical for a non-linear design process~(\textbf{R6}).

This case study of Nemo's workflow demonstrates how DashChat integrates multiple interaction modalities into a cohesive and efficient design experience. The system successfully guided the user from a vague initial concept to a detailed prototype through a continuous dialogue. By combining guided interactions via suggestion bubbles and interactive tags with the flexibility of natural language commands, DashChat significantly streamlines the generation and modification of dashboard prototypes. The History Panel further enhances this process by providing a safety net and a clear record of the design's evolution, ultimately transforming a traditionally fragmented process into a fluid and iterative conversation between the designer and the system.

%% file: sections/5evaluation.tex
\section{Evaluation}
In this section, we conduct a quantitative experiment to evaluate our pipeline and a user study to assess the usability.  
\label{sec:evalution}
\subsection{Pipeline Evaluation}
The evaluation quantitatively measures the accuracy and robustness of the DashChat pipeline, with a specific highlight on system Accuracy and hallucination mitigation. Our primary goal is to demonstrate the value of our multi-agent architecture compared to simpler, more direct LLM-based approaches.
\subsubsection{Experiment Setup}
We utilized a manually curated test set of 50 prompts, which is distinct from any examples used in few-shot prompting mentioned in \cref{sec:pipeline}, to evaluate the performance of DashChat and the baseline methods. The test set comprises 50 standard prompts covering various assigned intent tasks. For each prompt, a corresponding ground truth DSL output was manually authored by a human expert. We compare our full DashChat pipeline against two baselines:

 \begin{itemize}
    \item \textbf{Zero-Shot LLM:} A large language model~(Qwen2.5-Max) prompted with only the task description and DSL format, without our structured prompting, RAG, or iterative pipeline. This baseline isolates the contribution of our agentic architecture.
    \item \textbf{Tool Call LLM:} A baseline where the LLM~(Qwen2.5-Max) utilizes a single, monolithic function calling API to populate all DSL attributes. The task for LLM is to parse the prompt requirement from users and fill in the arguments for this single tool call. This baseline represents a more advanced but still non-agentic approach.
\end{itemize}

\subsubsection{Evaluation Metrics}
To provide a comprehensive assessment, we measured metrics across three categories: viability, consistency, and efficiency.

\textbf{Viability:} We first report the Execution Success Rate, as the measure of viability. This metric measures the percentage of generated DSL specifications that can be successfully parsed and rendered without any runtime errors. This is a standard practice in evaluating code generation systems~\cite{Hans2024On} and serves as a prerequisite for further consistency analysis.

\textbf{Consistency:} We measured Consistency to assess strict accuracy of standard prompt generation. A generated result is defined as consistent if it is identical to the ground truth specification across all key design attributes, including the chart types, their detailed axis characteristics and a detailed layout structure or color pattern suggestions. This metric serves as the most stringent measure of correctness, rewarding only those outputs that perfectly match the single, expert-authored ground truth. While it provides a clear signal of precision, its strictness may not fully account for the inherent ambiguity in natural language.

\textbf{Efficiency}: To directly evaluate our design requirement for fast prototyping, we measured Efficiency. This was quantified as the generation speed, defined as the end-to-end wall-clock latency in seconds, from prompt submission to the return of the final validated DSL specification.

\subsubsection{Results}
\begin{figure}[ht]
\centering
\includegraphics[width=\linewidth]{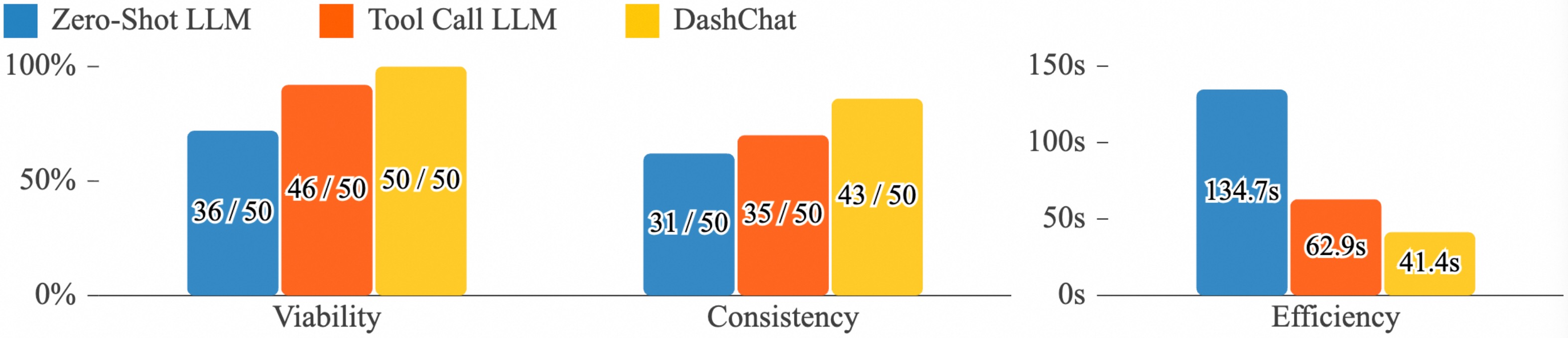}
\caption{The evaluation result shows the performance of DashChat, Zero-Shot LLM, and Tool Call LLM on different metrics.}
\label{fig:modelEval}
\end{figure}

Our evaluation results are presented in Fig.~\ref{fig:modelEval}, which indicates the performance of DashChat against the Zero-Shot LLM and Single Tool Call LLM baselines across our three metrics. The results indicate that DashChat significantly outperforms both baseline approaches in terms of viability, consistency, and efficiency. Notably, DashChat achieved perfect viability (100\%), the highest consistency (94\%), and the fastest generation speed (41.4s), demonstrating the effectiveness of our agentic architecture.

\textbf{Comparison with the Baselines}. Looking through the tested cases, two key factors stemming from our system's architecture account for the performance differences. The first is superior semantic understanding and reasoning. Unlike the single-pass baselines, our multi-agent pipeline decomposes the problem. The structured intent parsing and the agent planning RAG-infused process enable a more robust interpretation of user utterances, especially those with abstract or omitted information (\eg inferring an appropriate chart type from a task description). This directly contributes to DashChat's higher consistency score (94\%) compared to the Tool Call LLM (80\%) and the Zero-Shot LLM (76\%). The second factor is the explicit validation loop. DashChat's 100\% viability rate (50/50 successful executions) is a direct result of its Evaluation agent, which catches and corrects syntactically invalid or unexecutable specifications before finalizing the output. The baselines, lacking this mechanism, frequently failed, with the Zero-Shot LLM's viability dropping to as low as 72\%.

\textbf{Metric Difference and Efficiency Analysis}. An important observation is the relationship between the metrics. For DashChat, the consistency metric (94\%) is slightly lower than its perfect viability score (100\%), a difference that can be attributed to two main factors. First, ambiguity in abstract utterances often results in multiple reasonable answers. For instance, a prompt like "show sales distribution" might have a bar chart as its ground truth, but a generated pie chart would still be a viable and functionally correct response, thus passing the viability check but failing the stricter consistency test. Second, partially correct inferences occur when the model correctly generates the core visualization but misses a subtle, non-critical detail from the ground truth, such as a specific sorting order or a minor wording difference in a title. In terms of efficiency, DashChat's superior generation speed (41.4s) compared to the Tool Call (62.9s) and especially the Zero-Shot LLM (134.7s) may seem counterintuitive for a multi-step pipeline. We attribute this to the fact that the baselines' single, monolithic prompts are significantly more complex and computationally demanding for the LLM to process in one go. In contrast, DashChat's smaller, focused prompts for each agent are individually more efficient, and the overall orchestration results in a faster end-to-end response time.

\subsection{User Study} 
To comprehensively evaluate our system, we conducted a user study involving two distinct groups of participants: domain professionals and designers. It allowed us to assess both the system's effectiveness in producing contextually relevant prototypes and usability as an authoring tool.

\subsubsection{Participants}
We recruited qualified experts by sending invitations via email and advertised the study on social media platforms for additional participants. 
Our user study adhered to ethical standards, ensuring informed consent and data privacy while paying special attention to participants’ well-being.

\textbf{Group of Professionals (P)}. A total of 17 participants aged between 21 and 36 years old ($M = 26.12,~SD = 3.35$) were recruited, representing a diverse range of professional backgrounds~(denoted as P1-P17).
Among them, 6 identified themselves as male and 11 as female.
Their background included applied statistics~(3), biomedical engineering~(2), finance~(1), linguistics~(1), communication engineering~(2), manufacture~(2), aviation~(1), microelectronics~(2), and international politics~(3).
They are highly experienced in their respective domains with over two years of work experience ($M = 4.12,~SD = 2.18$) and possess a certain level of knowledge in data visualization.

\textbf{Group of Designers (D)}. We recruited 11 experienced designers~(7 females and 4 males), our target users, to operate our system~(D1-D11).
The participants ranged in age from 24 to 41 ($M = 31.18,~SD = 5.23$) and had more than two years of professional experience ($M = 6.73,~SD = 4.49$).
7 participants with prior experience in data visualization authoring tools (\eg Tableau~\cite{Tableau:2023:Learn}, Flourish~\cite{Flourish:2023:Flourish}) and 4 without any relevant experience.

\subsubsection{Procedure}
We conducted an in-person experiment for each participant. The study began with a 15-minute session where we introduced the system's concept, functionalities, and interaction methods. Following the introduction, participants engaged in tasks tailored to their respective groups:

Tasks for Group P: Participants were asked to propose a set of requirements for a performance dashboard based on their own domain knowledge. They then used our system to generate a prototype by providing textual prompts. The process was iterative, simulating a real-world negotiation scenario where they provided feedback to refine the prototype until they were satisfied. Their primary goal was to achieve a satisfactory and useful visual dashboard for their domain needs.

Tasks for Group D: Participants were given a series of structured tasks. The structured tasks required them to (1) generate a prototype for a given scenario~(\eg tobacco industry supply chain), and (2) systematically modify a prototype's views, layout, and style using both text and button interactions. In addition, they were asked to perform a within-study on the usability and effectiveness of the system against a state-of-the-art commercial tool, Tableau~\cite{Tableau:2023:Learn}. After a 30-minute tutorial on Tableau, each designer was asked to complete the same task of creating a dashboard prototype. The order of the tools was counterbalanced to mitigate learning effects. 

After completing their tasks, all participants were required to complete the questionnaire and participate in a semi-structured interview to provide detailed feedback.

\subsubsection{Measurements}

Our evaluation aims to capture a holistic view of the user experience. A unified questionnaire~(Q1-6), administered to all 28 participants, was first used to quantitatively assess the system against six fundamental metrics on a 5-point Likert scale (1 for ``strongly disagree'', 5 for ``strongly agree''). Especially, the designers need to assess both DashChat and Tableau:

\textbf{Usefulness}: The degree to which the system is perceived as being able to help users achieve their goals.

\textbf{Effectiveness}: The accuracy and completeness with which users can achieve their goals.

\textbf{Satisfaction}: The overall contentment and pleasantness of the user experience.

\textbf{Desirability}: The aesthetic appeal and attractiveness of the system and its output.

\textbf{Ease of Use}: The degree to which the system can be used without unnecessary effort.

\textbf{Ease of Learning}: The ease with which users can become proficient with the system.

While these core metrics provided a common baseline for all participants, our evaluation was further differentiated based on user type through tailored questions. The details of our questionnaire can be found in the supplementary material.

\textbf{For Group P}: The evaluation centered on the outcome and communicative value of the generated prototypes~(metrics for prototype generation). We focused on how well the system understood domain-specific requirements and translated them into meaningful visuals (related to \textbf{R2}, \textbf{R3}, \textbf{R4}). 6 interview questions~(Q7-12) probed the quality of data simulation, the relevance of chart choices, and the prototype's potential to accelerate communication in their professional context.

\textbf{For Group D}: The evaluation centered on the authoring process and creative workflow~(metrics for interface usability). We focused on the system's efficiency, flexibility, and its integration into a designer's toolkit (related to \textbf{R1}, \textbf{R5}, \textbf{R6}). 6 interview questions~(Q13-18) probed the efficiency of the text-to-prototype pipeline, the intuitiveness of the hybrid interaction model, and how the tool compared to traditional design software.

\subsubsection{Result and Findings}
\begin{figure*}[ht]
\centering
\includegraphics[width=\linewidth]{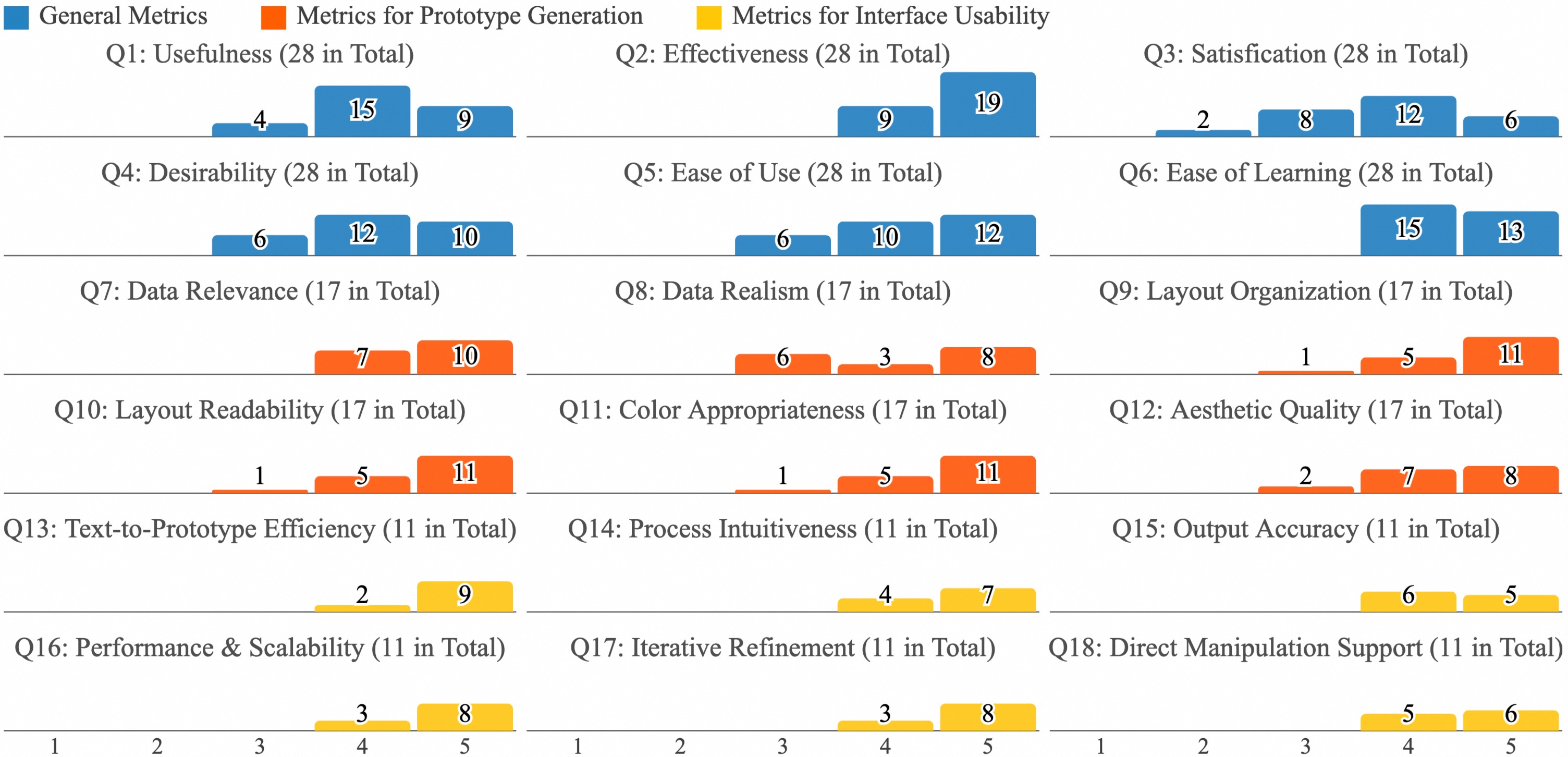}
\caption{Distribution of users' feedback on the questionnaire.}
\label{fig:user}
\end{figure*}

\begin{figure}[ht]
\centering
\includegraphics[width=\linewidth]{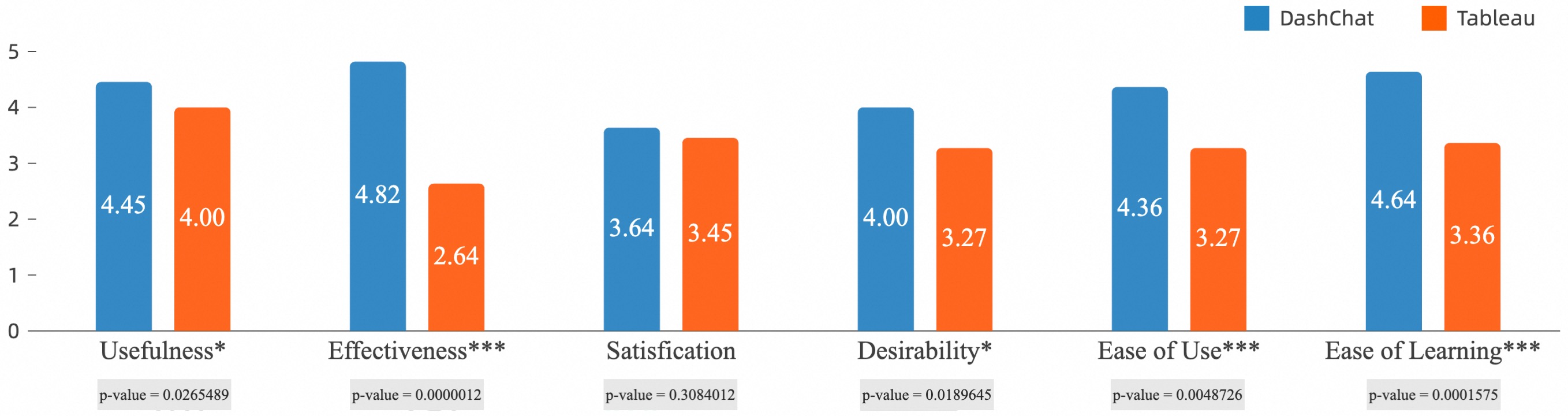}
\caption{Means of each measurement in DashChat and Tableau.}
\label{fig:within}
\end{figure}

Fig.~\ref{fig:user} and Fig.~\ref{fig:within} depict the quantitative results of the questionnaire, which indicate that our system was regarded favorably by participants.
We organize the feedback in detail.

\textbf{Overall Positive User Experience (N=28).}
Across all 28 participants, the system scored highly on the first six metrics, confirming its usability. It received exceptionally high ratings for Effectiveness ($M=4.68, SD=0.47$) and Ease of Learning ($M=4.46, SD=0.51$), indicating that users not only successfully achieved their goals but also found the system quick to master. Usefulness ($M=4.18, SD=0.72$), Ease of Use ($M=4.21, SD=0.80$), and Desirability ($M=4.14, SD=0.76$) also received strong positive feedback, demonstrating that participants found the tool valuable, effortless to operate, and aesthetically pleasing. The overall Satisfaction ($M=3.79, SD=0.96$) was also positive, though exhibiting slightly more variance, reflecting the nuanced and specific feedback provided by different user groups.

\textbf{Superior Usability and Effectiveness}. As shown in Fig.~\ref{fig:within}, a paired t-test revealed that DashChat received significantly higher scores than Tableau. The advantages were most pronounced for Effectiveness ($M=4.82$ vs. $M=2.64$; $p < .001$), Ease of Learning ($M=4.64$ vs. $M=3.36$; $p < .001$), and Ease of Use ($M=4.36$ vs. $M=3.27$; $p < .01$). DashChat also scored significantly higher on Desirability ($p < .05$) and Usefulness ($p < .05$). No significant difference was found for Satisfaction ($p = .31$), suggesting that while both tools can produce a satisfactory final product, the process to get there is markedly different. The interview feedback clearly explained the result. D5 noted, \textit{``With DashChat, I described the dashboard and got a full, multi-chart layout in under a minute. With Tableau, I had to start from a blank canvas, manually drag each data field, and build each chart one by one.''} This directly explains the dramatic difference in Effectiveness. Similarly, D8 highlighted the gentler learning curve: \textit{``Tableau forced me to think about dimensions versus measures. DashChat just understood 'sales trends' and did the right thing. It felt like a conversation, not a technical task,''} which accounts for the higher Ease of Use and Ease of Learning scores.

\textbf{High-quality generated prototypes.}
We found that participants were generally positive about the generated prototypes.
First, participants expressed general satisfaction with the chart generation~(\textbf{R2}).
P9 commented that \textit{``the simulated data was surprisingly realistic.''}
P9, from the manufacturing domain, requested a prototype focused on production efficiency across factories.
In response, the system automatically simulated output data for morning, afternoon, and night shifts at each factory, information not explicitly mentioned in the input. 
\textit{``This shift-based data is exactly what we monitor in our daily operations,''} indicating that the system's proactive data simulation aligned closely with real-world industry practices and user expectations.
However, P3 and P17 felt this functionality could be further improved, particularly when LLMs failed to interpret certain domain-specific acronyms correctly. 
P17 further suggested adding the web search to enhance data simulation accuracy and relevance.
Moreover, reasonable chart-type selections greatly supported effective data communication.
P16 remarked, \textit{``I was surprised that the charts matched what I had in mind. It used a line chart to show trends over time, saving me from explaining every detail.''}

Regarding layout~(\textbf{R3}), most participants appreciated the system to organize multiple views coherently and informally. 
P6 stated, \textit{``The layout looks clean and pleasing, balancing visual density and whitespace.''}
P2, P3, and P11 favored how the system automatically prioritized key views based on their intent.
They remarked that \textit{``the layout of the dashboard views was well-organized, allowing me to quickly locate the key metrics without visual clutter.''}
P11, a communication engineering professional, requested emphasis on base station counts across Chinese provinces. 
The system responded with a geographic heatmap, centrally positioned and occupying the largest screen area.
P15 suggested improving the layout by placing semantically related views closer to enhance interpretation and comparison.
Aesthetically, the visual style was also well received~(\textbf{R4}), with P8 and P9 commenting that \textit{``the design looks professional and polished, and the color scheme enhances readability without being overwhelming.''}
The incorporated stylistic elements enhanced the aesthetic sophistication, \textit{``I liked how the system added subtle design touches, like borders, icons, and animations.''}

\textbf{Accelerated negotiation process.}
Many participants emphasized that the system substantially enhances communication efficiency, particularly in the early design stages.
\textit{``Seeing the visuals update instantly based on my feedback makes me feel like we were all on the same page much faster.''}
By quickly transforming high-level textual requirements into concrete visual prototypes, our system served as a shared reference point that reduced ambiguity and accelerated the alignment of expectations.
P9 remarked, \textit{``It helped me communicate my ideas without needing to draw anything or explain every detail. I could just react to what I saw.''} 
P6 added, \textit{``Normally, it takes multiple meetings to get the visual right, but with this system, we aligned in just a few iterations.''}

\textbf{Effective and efficient authoring process.}
Participants consistently praised the system for significantly streamlining the construction of prototypes, reducing both design complexity and turnaround time~(\textbf{R1}, \textbf{R5}). 
D2 and D5 noted, \textit{``Compared to traditional performance dashboard construction, the 1-minute generation speed is significantly faster.''}, illustrating a substantial speed improvement compared to traditional methods.
D6 also expressed satisfaction, \textit{``It allows me to quickly build a prototype, enabling face-to-face communication with clients to showcase design details. If clients request sudden changes, the system responds promptly with updated outputs, streamlining communication and reducing turnaround time.''}
It also aligns with the findings from our previous user study.
Besides that, participants are positive that the system bridges the gap between abstract textual input and concrete visual output.
D7 commented, \textit{``The generated prototype closely aligned with the textual requirements, with minimal discrepancies.''}
The simulated data functionality was appreciated by participants, \textit{``it helps designers without a strong data background to save the time that would otherwise be spent creating placeholder data manually.''}.
Additionally, D8 noted that the prototypes demonstrated a high level of visual fidelity, including well-balanced layouts, appropriate color schemes, and realistic chart compositions, \textit{``The generated output closely resembles real-world performance dashboards in terms of visual appearance''}

\textbf{User-friendly interactions.}
Our system received positive feedback for its interactive flexibility, particularly in supporting both iterative refinement and direct manipulation~(\textbf{R6}).
The combination of the Q\&A system and in-bubble button allows users to freely choose their preferred interaction mode, enhancing overall flexibility and user autonomy.
D4 commented, \textit{``The chat bubble interaction design is particularly appealing, as it offers a more efficient way to operate.''}
The ability to provide shortcuts for common modifications was also highly valued.
D3 and D6 noted that shortcuts facilitate faster text input and additionally serve as a source of creative inspiration.
Additionally, some participants provided detailed suggestions for refining our interaction design.
D4 suggested, \textit{``Sometimes I just want to tweak one color without changing the whole palette. Having that option would make the system feel more customizable.''}
D9 and D10 suggested, \textit{``Direct on-prototype editing would further improve usability.''}

%% file: sections/6discussion.tex
\section{Discussion}
\textbf{Implications and generalization of DashChat.}
Our work demonstrates how multi-agent interaction powered by LLMs can be effectively harnessed to support performance dashboard authoring, a task that traditionally requires both domain expertise and design proficiency. 
DashChat supports a more inclusive and collaborative design process between humans and AI by enabling users to construct high-quality dashboard prototypes through natural language.
By distributing sub-tasks across multiple specialized agents running in parallel, our pipeline offers a new approach to improving the responsiveness of LLM-powered multi-agent workflows.
This parallelism mitigates the latency commonly associated with sequential LLM pipelines, which provides a scalable solution for complex generation tasks.
Beyond the domain, our system's underlying architecture and interaction paradigm exhibit strong potential for generalization across broader domains of LLM-assisted visual design.
We structure the design space as a knowledge base to guide LLM-based generation, define task-specific automation pipelines based on user intent, and design interaction mechanisms that support user-centered refinement.
The methodology offers a generalizable blueprint for combining LLMs with domain knowledge and interaction design, applicable to a wide range of visual communication tasks.

\textbf{Creativity and robustness of LLMs.}
LLMs are powerful tools for bridging natural language and visual representations, allowing even non-experts to generate meaningful visualizations through text-based interaction.
This capability lowers the barrier to entry for data storytelling and democratizes access to visualization tools by translating human intent into structured visual output.
However, the application of LLMs in visualization reveals a fundamental tension between creativity~(generative diversity) and robustness~(output stability).
Emphasizing creativity allows for diverse and potentially novel visual representations, opening new avenues for expressive and personalized design, but unconstrained generation may violate core visualization design principles.
Conversely, overly deterministic strategies~(\eg relying solely on fixed templates) may preserve correctness but suppress innovation, which results in homogeneous outputs that lack adaptability to specific user contexts or domain narratives.
Because LLMs generate visual descriptions based on probabilistic language patterns rather than grounded reasoning or formal semantics, they often lack the logical rigor required in data visualization. 
Effective generation must integrate knowledge from data science, visual design, and domain expertise.
In this work, we constrained the generative space with a validated knowledge base, leveraging trusted templates and refining them based on user preferences.
The conflict can be mitigated by designing such hybrid frameworks carefully in the short term.
Resolving this tension in the long term may require paradigm-level shifts of LLMs, such as unifying statistical learning with logical reasoning~\cite{havrilla2024glorewhenwhereimprove, wang2024rethinkingboundsllmreasoning}.

\textbf{Limitations and future work.}
First, our system has room for improvement in task composition and data simulation.
The first user study shows this limitation may cause mismatches between simulated data and user intent, reducing the output's utility and credibility.
The ability could be enhanced by integrating external knowledge sources via techniques such as web search or RAG~\cite{2023arXiv231210997G}.
These approaches enable LLMs to access up-to-date, domain-specific knowledge beyond their pre-trained parameters, allowing for more accurate and context-aware responses.
Second, we mainly focus on the global palette recommendation and embellishment generation centering on a color-related design style. 
Broader design dimensions, such as typography and iconography, should be considered to enhance stylistic richness.
Incorporating the additional elements would allow for richer stylistic expression, better alignment with user preferences and domain contexts, and greater adaptability across different application scenarios.
\section{Conclusion}
\label{sec:discussion}
This work presents DashChat,  an interactive system that leverages LLMs to generate performance dashboard design prototypes from natural language.
We first summarized their common design patterns by analyzing high-quality performance dashboards.
Next, we built a multi-agent pipeline powered by LLMs to generate practical, aesthetic prototypes.
Then, we developed a user-friendly interface that supports text-based interaction.
Last, we evaluated DashChat via a quantitative experiment and a user study, which showed that DashChat can generate high-quality prototypes rapidly while offering a user-friendly interactive experience.